\documentclass[acmtog]{acmart}
\pdfoutput=1
\usepackage[utf8]{inputenc}
\usepackage[english]{babel}
\usepackage{amsmath}
\usepackage{amsfonts}
\usepackage{placeins}
\usepackage{amssymb}
\usepackage{hyperref}
\usepackage{graphbox}
\usepackage[normalem]{ulem}
\usepackage{rotating}
\setcitestyle{square} 

\author{Valentin Deschaintre} \affiliation{\institution{Inria, Optis, Universit\'{e} C\^{o}te d'Azur}} \email{valentin.deschaintre@inria.fr} 
\author{Miika Aittala} \affiliation{\institution{MIT CSAIL}} 
\author{Fredo Durand} \affiliation{\institution{MIT CSAIL, Inria, Universit\'{e} C\^{o}te d'Azur}} 
 \author{George Drettakis} \affiliation{\institution{Inria, Universit\'{e} C\^{o}te d'Azur}} 
\author{Adrien Bousseau} \affiliation{\institution{Inria, Universit\'{e} C\^{o}te d'Azur}} 

\definecolor{todoCol}{rgb}{1,0.5,0}
\newcommand{\NEW}  [1] {{#1}}
\newcommand{\REM}  [1] {}

\setcitestyle{square}
\title{Single-Image SVBRDF Capture with a Rendering-Aware Deep Network}
\begin{document}

\begin{abstract}
Texture, highlights, and shading are some of many visual cues that allow humans to perceive material appearance in single pictures.
Yet, recovering spatially-varying bi-directional reflectance distribution functions (SVBRDFs) from a single image based on such cues has challenged researchers in computer graphics for decades.
We tackle lightweight appearance capture by training a deep neural network to automatically extract and make sense of these visual cues.
Once trained, our network is capable of recovering per-pixel normal, diffuse albedo, specular albedo and specular roughness from a single picture of a flat surface lit by a hand-held flash. 
We achieve this goal by introducing several innovations on training data acquisition and network design.
For training, we leverage a large dataset of artist-created, procedural SVBRDFs which we sample and render under multiple lighting directions. We further amplify the data by material mixing to cover a wide diversity of shading effects, which allows our network to work across many material classes. Motivated by the observation that distant regions of a material sample often offer complementary visual cues, we design a network that 
combines an encoder-decoder convolutional track for local feature extraction with a fully-connected track for \emph{global feature} extraction and propagation.
Many important material effects are view-dependent, and as such ambiguous when observed in a single image. We tackle this challenge by defining the loss as a differentiable SVBRDF similarity metric that compares the \emph{renderings} of the predicted maps against renderings of the ground truth from several lighting and viewing directions.
Combined together, these novel ingredients bring clear improvement over state of the art methods for single-shot capture of spatially varying BRDFs.
\\\\
\textit{This is the author's version of the work. It is posted here for your personal use. Not for redistribution. The definitive version was published in ACM Trans. Graph. 37, 4, Article 128 (August 2018), https://doi.org/10.1145/3197517.3201378.}
\textcolor{red}{This version was compressed for archive, a higher quality version is available on the project web page: \url{https://team.inria.fr/graphdeco/fr/projects/deep-materials/ .}}
\end{abstract}



\setcopyright{None}
\acmJournal{TOG}
\acmYear{2018}\acmVolume{37}\acmNumber{4}\acmArticle{128}\acmMonth{8} \acmDOI{10.1145/3197517.3201378}

\begin{teaserfigure}
\includegraphics[width=\textwidth]{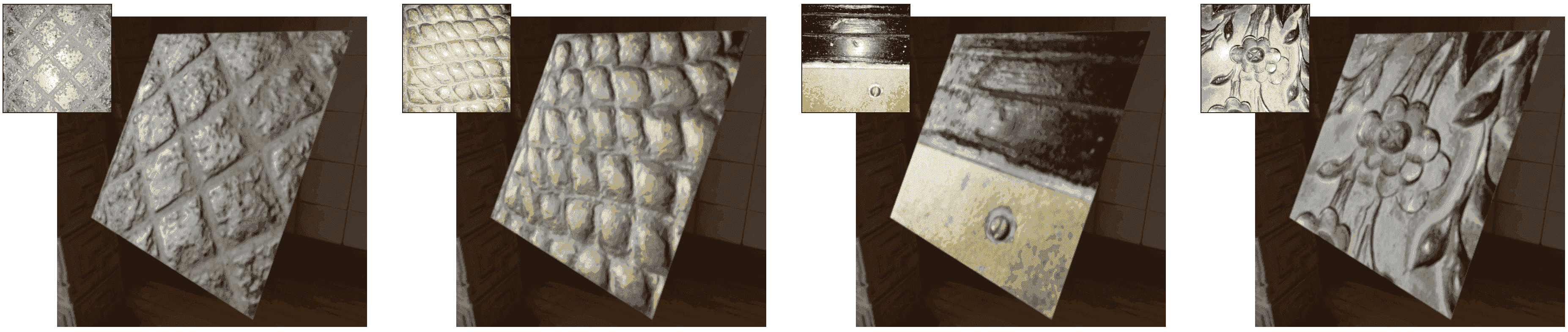}
\vspace{-5mm}
\caption{From a single flash photograph of a material sample (insets), our deep learning approach predicts a spatially-varying BRDF. See supplemental materials for animations with a moving light.}
\label{fig:teaser}
\end{teaserfigure}

\maketitle

\section{Introduction}

The appearance of real-world objects results from complex interactions between light, reflectance, and geometry.
Disentangling these interactions is at the heart of \emph{lightweight appearance capture}, which aims at recovering reflectance functions from one or a few photographs of a surface.
This task is inherently ill-posed, since many different reflectances can yield the same observed image. For example, any photograph can be perfectly reproduced by a diffuse albedo map, where highlights are ``painted'' over the surface.

A combination of two strategies is generally employed to deal with this ill-posedness. First, ambiguity can be reduced by collecting additional measurements under different viewing or lighting conditions. 
While this strategy is currently the most appropriate to achieve high accuracy, it requires precise control of the acquisition process \cite{Xu2016}.
The second strategy is to introduce \emph{a priori} assumptions about the space of plausible solutions. 
While designing such priors by hand has challenged researchers for decades \cite{Guarnera16}, Convolutional Neural Networks (CNNs) have emerged as a powerful method to automatically \emph{learn} effective priors from data. 

In this paper, we propose a deep learning approach to single-image appearance capture, where we use \emph{forward} rendering simulations 
to train a neural network to solve the ill-posed \emph{inverse} problem of estimating a spatially-varying \NEW{bi-directional} reflectance \NEW{distribution function} (SVBRDF) from one picture of a flat surface lit by a hand-held flash. While our method shares ingredients with recent work on material capture \cite{rematas2017,Li17}, material editing \cite{liu2017material}, and other image-to-image translation tasks \cite{isola17}, achieving high-quality SVBRDF estimation requires several key innovations on training data acquisition and neural network design.


A common challenge in supervised learning is the need for many training images and the corresponding solutions. For materials, this problem is acute: even with the most lightweight capture methods, we cannot obtain enough measurements to train a CNN. Furthermore, such an approach would inherit the limitations of the data capture methods themselves. Following the success of using synthetic data for training \cite{Su15,Zhang16,Richter2016}, we tackle this challenge by leveraging a large dataset of artist-created, procedural SVBRDFs \cite{Substance}, which we sample and render under multiple lighting directions to create training images. We further amplify the data by randomly mixing these SVBRDFs together and render multiple randomly scaled, rotated and lit versions of each material, yielding a training set of up to $200{,}000$ realistic material samples. 

The task of our deep network is to predict four maps corresponding to \emph{per-pixel} normal, diffuse albedo, specular albedo, and specular roughness of a planar material sample. 
However, directly minimizing the pixel-wise difference between predicted and ground truth parameter maps is suboptimal, as it does not consider the interactions between variables. Intuitively, while a predicted map may look plausible when observed in isolation, it may yield an image far from the ground truth when combined with other maps by evaluating the BRDF function.
Furthermore, the numerical differences in the parameter maps might not consistently correlate with differences in the material's appearance, causing a naive loss to weight the importance of different features arbitrarily.
We mitigate these shortcomings by formulating a differentiable SVBRDF similarity metric that compares the \emph{renderings} of the predicted maps against renderings of the ground truth from several lighting and viewing directions. 


We focus on lightweight capture by taking as input a single near-field flash-lit photograph. \REM{Such images}\NEW{Flash photographs} are easy to acquire, and have been shown to contain a lot of information that can be leveraged in inferring the material properties \NEW{from one} \cite{Ashikhmin07,Aittala15,Aittala16} \NEW{or multiple images \cite{Riviere2016,Hui2017}}. In such images, the pixels showing the highlight provide strong cues about specularity, whereas the outer pixels show diffuse and normal variations more prominently. To arrive at a consistent solution across the image, these regions need to share information about their respective observations. 
Unfortunately, our experiments reveal that existing encoder-decoder architectures struggle to aggregate distant information and propagate it to fine-scale details. To address this limitation, we enrich such architectures with a secondary network that extracts global features at each stage of the network and combines them with the local activations of the next layer, facilitating back-and-forth exchange of information across distant image regions.

\begin{figure}[!t]
\includegraphics[width=\linewidth]{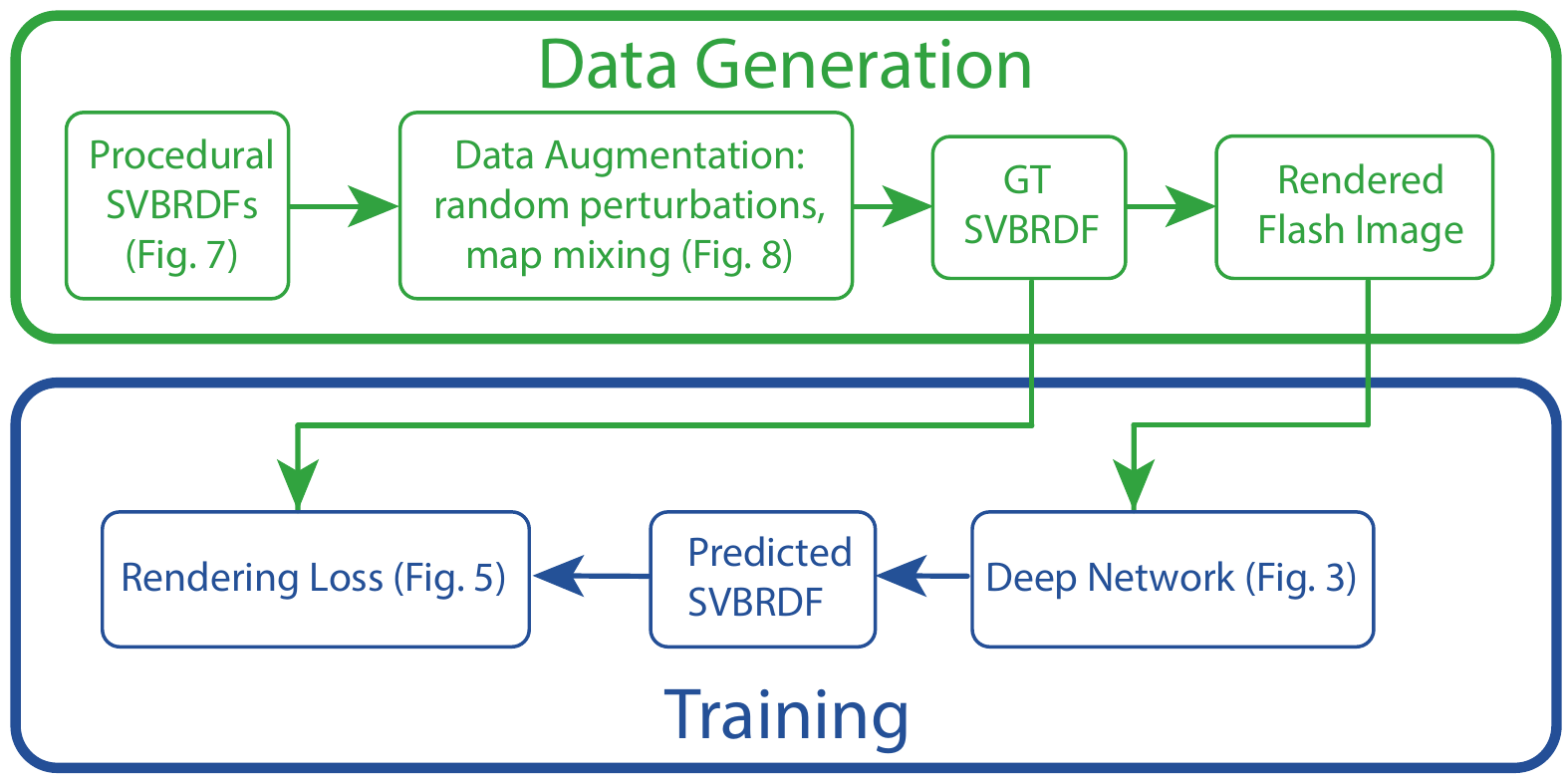}
\caption{\NEW{Overview of our method: we use procedural SVBRDFs to generate our ground truth (GT) training data, which we augment by random perturbations of the procedural parameters and mixing of the SVBRDF maps (Fig.~\ref{fig:SubstanceDiversity} and~\ref{fig:dataAugmentation}, Sec.~\ref{sec:training_data}). We then use physically-based rendering to synthesize the corresponding flash images. These are used to train our Deep Network (Fig.~\ref{fig:overviewArchi}, Sec.~\ref{sec:unet} and~\ref{sec:gnet}) which compares predicted SVBRDFs and ground truth using a rendering loss (Fig.~\ref{fig:renderLoss}, Sec.~\ref{sec:renderloss}).  }
}
\label{fig:overviewFlow}
\end{figure}

In summary, we introduce a method to recover spatially-varying diffuse, specular and normal maps from a single image captured under flash lighting. Our approach outperforms existing work \cite{Li17,Aittala16} on a wide range of materials thanks to several technical contributions \NEW{(Fig.~\ref{fig:overviewFlow})}:
\begin{itemize}
\item We exploit procedural modeling and image synthesis to generate a very large number of realistic SVBRDFs for training. We \NEW{provide} this dataset freely for research purposes\footnote{\NEW{\url{https://team.inria.fr/graphdeco/projects/deep-materials/}}}. 
\item We introduce a \emph{rendering loss} that evaluates how well a prediction reproduces the appearance of a ground-truth material sample.
\item We introduce a secondary network to combine global information extracted from distant pixels with local information necessary for detail synthesis.
\end{itemize}

\NEW{We stress that our goal is to approximate the appearance of a casually-captured material rather than recover accurate measurements of its constituent maps.}



\section{Related Work}
The recent survey by Guarnera et al. \cite{Guarnera16} provides a detailed discussion of the wide spectrum of methods for material capture. Here we focus on lightweight methods for easy capture of spatially-varying materials in the wild.

A number of assumptions have been proposed to reduce ambiguity when only a few measurements of the material are available.
Common priors include spatial and angular homogeneity \cite{Zickler06}, repetitive or random texture-like behavior \cite{Wang2011,Aittala15,Aittala16}, sparse environment lighting \cite{Lombardi16,Dong14}, polarization of sky lighting \cite{Riviere17}, mixture of basis BRDFs \cite{Ren11,Dong10,Hui2017}, optimal sampling directions \cite{Xu2016}, and user-provided constraints \cite{AppGen2011}. However, many of these assumptions restrict the family of materials that can be captured. For example, while the method by Aittala et al.~\shortcite{Aittala16} takes a single flash image as input, it cannot deal with non-repetitive material samples (see Section \ref{sec:comparisons}). We depart from this family of work by adopting a data-driven approach, where a neural network learns its own internal assumptions to best capture the materials it is given for training.

Dror et al.~\shortcite{Dror2001} were among the first to show that a machine learning algorithm can be trained to classify materials from low-level image features. Since then, deep learning emerged as an effective solution to related problems such as intrinsic image decomposition \cite{Narihira2015,innamorati2017} and reflectance and illumination estimation \cite{rematas2017}. Most related to our approach is the work by Li et al.~\shortcite{Li17}, who adopted an encoder-decoder architecture similar to ours to estimate diffuse reflectance and normal maps. However, their method only recovers uniform specular parameters over the material sample. In contrast, we seek to recover per-pixel specular albedo and roughness. Furthermore, they trained separate networks for different types of materials, such as wood and plastic. Rather than imposing such a hard manual clustering (which is ambiguous anyway: consider the common case of plastic imitation of wood), we train a single all-purpose network and follow the philosophy of letting it learn by itself any special internal treatment of classes that it might find useful. 
In addition, Li et al.~\shortcite{Li17} introduce a strategy called \emph{self-augmentation} to expand a small synthetic training set with semi-synthetic data based on the network's own predictions for real-world photographs. This strategy is complementary to our massive procedural data generation.

Since our goal is to best reproduce the appearance of the captured material, we evaluate the quality of a prediction using a differentiable \emph{rendering loss}, which compares renderings of the predicted material with renderings of the ground truth given for training. 
Rendering losses have been recently introduced by Tewari et al.~\shortcite{tewari17MoFA} and Liu et al.~\shortcite{liu2017material} for facial capture and material editing respectively. Tewari et al. use a rendering loss to compare their reconstruction with the input image in an unsupervised manner, while Liu et al. use it to evaluate their reconstruction with respect to both the input image and a ground-truth edited image.
Aittala et al. \shortcite{Aittala16} also use a differentiable renderer to compare the textural statistics of their material estimates with those of an input photograph. However, they use this loss function within a standard inverse-rendering optimization rather than to train a neural network. In contrast to the aforementioned methods, our rendering loss is a \emph{similarity metric} between SVBRDFs. The renderings are not compared to the input photograph, as this would suffer from the usual ambiguities related to single-image capture. Rather, we compare the rendered appearance of the estimated and the target material under many lighting and viewing directions, randomized for each training sample. 

The need for combining local and global information appears in other image transformation tasks. In particular, Iizuka et al.~\shortcite{IizukaSIGGRAPH2016} observe that colors in a photograph depend both on local features, such as an object's texture, and global context, such as being indoor or outdoor. Based on this insight, they propose a convolutional network that colorizes a gray-level picture by separately extracting global semantic features and local image features, which are later combined and processed to produce a color image. \NEW{Contextual information also plays an important role in semantic segmentation, which motivates Zhao et al. \shortcite{ZhaoSQWJ17} to aggregate the last layer feature maps of a classification network in a multi-scale fashion.} While we also extract local and global features, we exchange information between these two tracks after every layer, allowing the network to repeatedly transmit information across all image regions.
Wang et al.~\shortcite{Wang17} introduced a related non-local layer that mixes features between all pixels, and can be inserted at multiple points in the network to provide opportunities for non-local information exchange. While they apply more complex nonlinear mixing operations, they do not maintain an evolving global state across layers. \NEW{Finally, \emph{ResNet} \cite{HeZRS16} aims at facilitating information flow between co-located features on different layers, which makes the training better behaved. Our architecture has a complementary goal of aiding efficient global coordination between non-co-located points. Our scheme also opens up novel pathways, allowing information to be directly transmitted between distant image regions.}

\begin{figure*}[th]
\includegraphics[width=\textwidth]{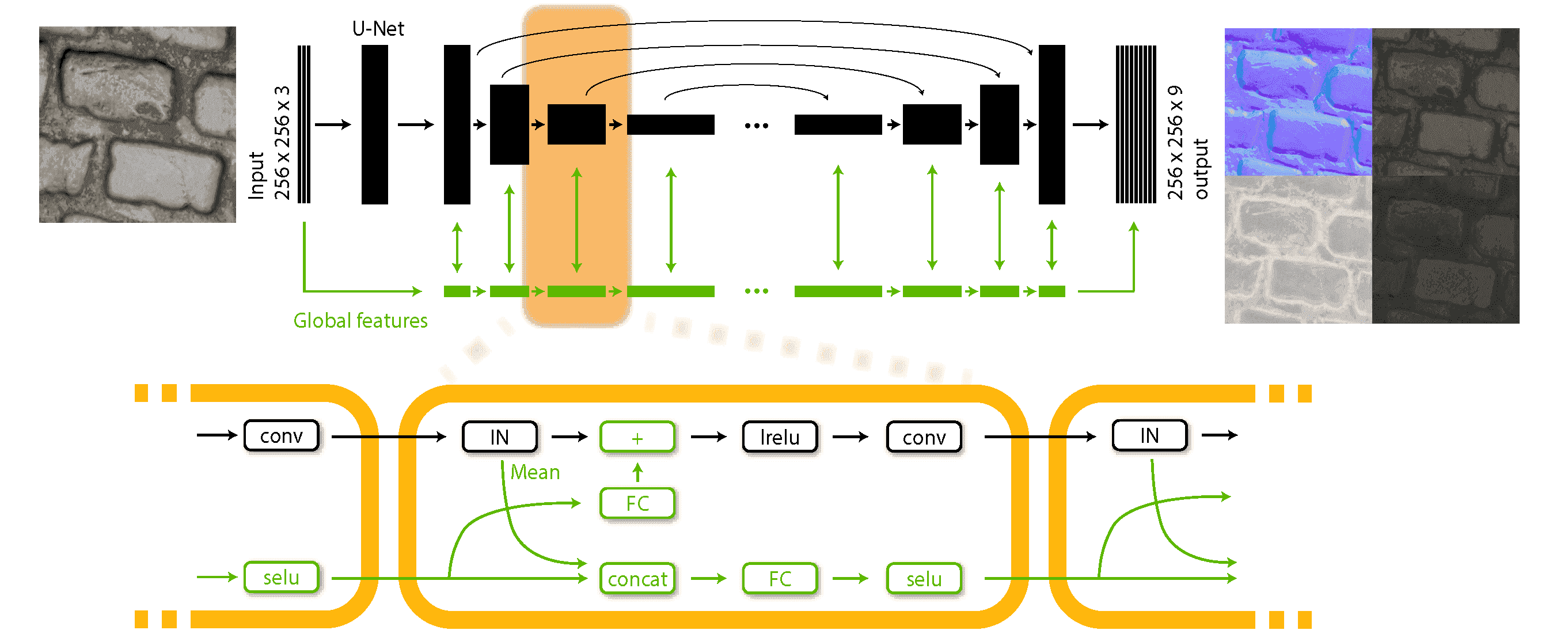}
\caption{\REM{Overview}\NEW{Architecture} of our deep convolutional network, which takes as input a single flash-lit image (left) and predicts four maps corresponding to per-pixel normal, diffuse albedo, specular albedo and specular roughness (right).
	Our network follows the popular U-Net encoder-decoder architecture (black), which we complement with a new \emph{global features} \NEW{track} (green) that processes vectors instead of feature maps. Taken together, the full network consists of repeating ``modules'', which are detailed in the bottom part of the figure.
	At every stage of the network, the feature means subtracted by the instance normalization after the convolutional layer are concatenated with the global feature vector, which is then processed by a fully connected layer and a non-linearity before being added to the feature maps of the next stage. \emph{IN} and \emph{FC} denote instance normalizations and fully connected layers respectively. We use SELU \cite{Klambauer17} and leaky ReLu activation functions. In the decoder, the set of layers also includes a skip-connection concatenation and a second convolution, which we omit for clarity. We \REM{will}provide the code of our network to allow reproduction.}
\label{fig:overviewArchi}
\end{figure*}

\section{Network Architecture}
Our problem boils down to translating a photograph of a material into a coinciding SVBRDF map representation, which is essentially a multi-channel image. The \emph{U-Net} architecture \cite{Ronneberger15} has proven to be well suited for a wide range of similar image-to-image translation tasks \cite{zhang2017real,isola17}. However, our early experiments revealed that despite its multi-scale design, this architecture remains challenged by tasks requiring the fusion of distant visual information. We address this limitation by complementing the U-Net with a parallel \emph{global features} network tailored to capture and propagate global information.
\subsection{U-Net Image-to-Image Network}
\label{sec:unet}
We adopt the U-Net architecture as the basis of our network design, and follow Isola et al.~\shortcite{isola17} for most implementation details. Note however that we do not use their \emph{discriminator} network, as we did not find it to yield a discernible benefit in our problem.
We now briefly describe the network design. We \REM{will}provide the code of our network and its learned weights to allow reproduction of our results\footnote{\NEW{\url{https://team.inria.fr/graphdeco/projects/deep-materials/}}}.

As illustrated in Figure~\ref{fig:overviewArchi}, our base network takes a $3$-channel photograph as input and outputs a $9$-channel image of SVBRDF parameters -- $3$ channels for the RGB diffuse albedo, $3$ channels for the RGB specular albedo, $2$ channels for the $x$ and $y$ components of the normal vector \NEW{in tangent plane parameterization}, and $1$ channel for the specular roughness.
We use low dynamic range images as input photographs due to the ease of acquisition, and let the network learn how to interpret the saturated highlight regions. Regardless, the dynamic range of flash photographs can still be large. We flatten the dynamic range by transforming the input image into logarithmic space and compacting it to the range $[0,1]$ via the formula $\frac{\log (x+0.01) - \log 0.01}{\log(1.01) - \log(0.01)}$.

The input image is processed through a sequence of $8$ convolutional layers that perform downsampling (the encoder), followed by a sequence of $8$ upsampling and convolutional layers (the decoder). Such a hourglass-shaped network gradually reduces the resolution of the image while increasing the feature size, forcing the encoder to compress the relevant information into a concise, global feature vector. The task of the decoder is to expand these global features back into a full-sized image that matches the training target. However, while the bottleneck is critical to aggregate spatially-distant information, it hinders the reproduction of fine details in the output. Following Ronneberger et al.~\shortcite{Ronneberger15}, we mitigate this issue by introducing skip connections between same-sized layers of the encoder and decoder, helping the decoder to synthesize details aligned with the input at each spatial scale. 

Prior to the decoder, we insert a single convolutional layer with 64 output feature channels. The feature counts in the encoder downscaling layers are 128, 256, 512, 512, 512, 512, 512 and 512. The downsampling is implemented by using a stride of $2$ in the convolutions. In the decoder, the same feature counts are used in reverse order. At each scale, a nearest-neighbor upsampling is followed by concatenation of encoder features, and two convolutions. We use the filter size $[4,4]$ across all layers. For nonlinearities we use the leaky ReLu activation function with a weight $0.2$ for the negative part. The final output is mapped through a sigmoid to enforce output values in the range $[0, 1]$.

Following each convolution layer (or pair thereof), we apply instance normalization, which stabilizes training on image generation tasks \cite{Ulyanov_2017_CVPR,isola17}. Finally, we regularize by applying dropout at $50\%$ probability on the three coarsest layers of the decoder. 

\subsection{Global Features Network} 
\label{sec:gnet}
Distant regions of a material sample often offer complementary information to each other for SVBRDF recovery. This observation is at the heart of many past methods for material capture, such as the work of Lensch et al.~\shortcite{Lensch2003} where the SVBRDF is assumed to be spanned by a small set of basis BRDFs, 
or the more recent work of Aittala et al. \shortcite{Aittala15, Aittala16} where spatial repetitions in the material sample are seen as multiple observations of a similar SVBRDF patch. 
Taking inspiration from these successful heuristics, we aim for a network architecture capable of leveraging \REM{any}redundancies present in the data.

\begin{figure}
\begin{tabular}{cccc}
\hspace{-3mm}\includegraphics[width=0.24\linewidth]{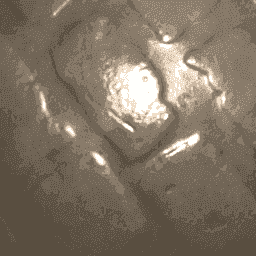} &
\hspace{-3mm}\includegraphics[width=0.24\linewidth]{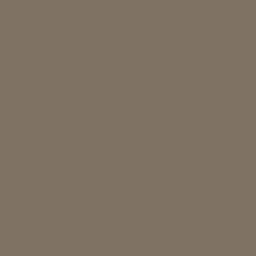} &
\hspace{-3mm}\includegraphics[width=0.24\linewidth]{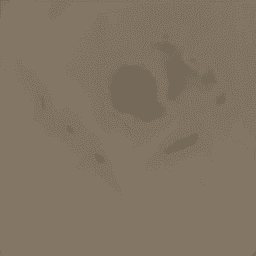} &
\hspace{-3mm}\includegraphics[width=0.24\linewidth]{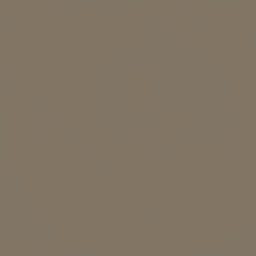} \\ 
\hspace{-3mm}\small{(a) Input} & 
\hspace{-3mm}\small{(b) GT Average} & 
\hspace{-3mm}\small{(c) U-Net} &
\hspace{-3mm}\small{(d) Ours} \\
\\

\hspace{-3mm} \hspace{13mm} \begin{sideways} \hspace{5mm} \small{Normal} \end{sideways} &
\hspace{-3mm}\includegraphics[width=0.24\linewidth]{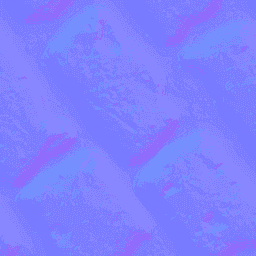} &
\hspace{-3mm}\includegraphics[width=0.24\linewidth]{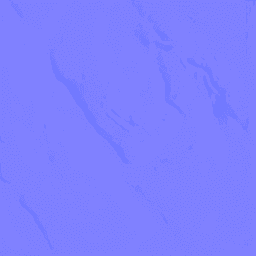} &
\hspace{-3mm}\includegraphics[width=0.24\linewidth]{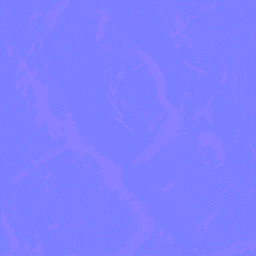} \\ 
 
\hspace{-3mm} \hspace{13mm} \begin{sideways} \hspace{1mm} \small{Diffuse albedo} \end{sideways} &
\hspace{-3mm}\includegraphics[width=0.24\linewidth]{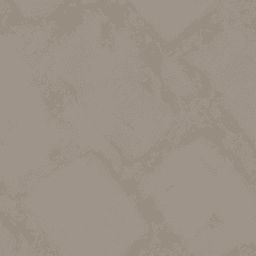} &
\hspace{-3mm}\includegraphics[width=0.24\linewidth]{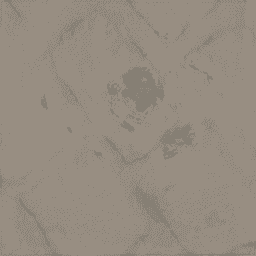} &
\hspace{-3mm}\includegraphics[width=0.24\linewidth]{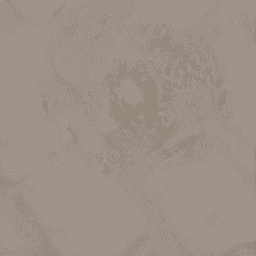} \\ 

\hspace{-3mm} \hspace{13mm} \begin{sideways} \hspace{4mm} \small{Roughness} \end{sideways} &
\hspace{-3mm}\includegraphics[width=0.24\linewidth]{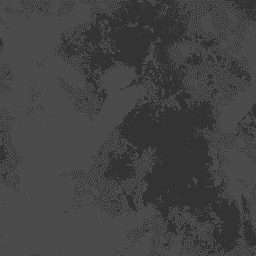} &
\hspace{-3mm}\includegraphics[width=0.24\linewidth]{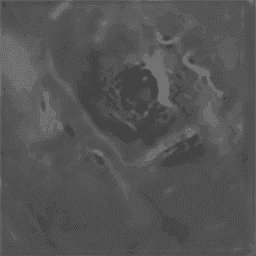} &
\hspace{-3mm}\includegraphics[width=0.24\linewidth]{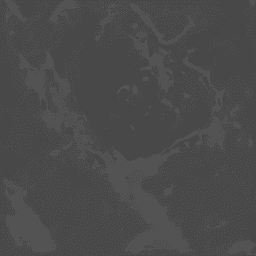} \\ 

\hspace{-3mm} \hspace{13mm} \begin{sideways} \hspace{.5mm} \small{Specular albedo} \end{sideways} &
\hspace{-3mm}\includegraphics[width=0.24\linewidth]{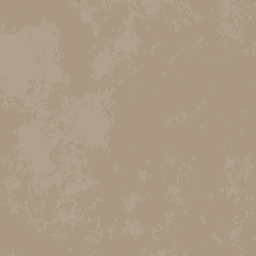} &
\hspace{-3mm}\includegraphics[width=0.24\linewidth]{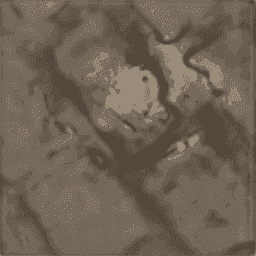} &
\hspace{-3mm}\includegraphics[width=0.24\linewidth]{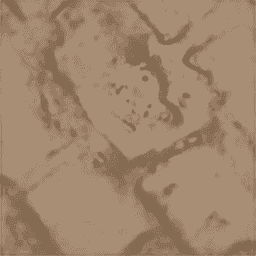} \\
\hspace{-3mm}& 
\hspace{-3mm}\small{(e) GT SVBRDF} & 
\hspace{-3mm}\small{(f) U-Net} &
\hspace{-3mm}\small{(g) Ours} 

\end{tabular} 
\caption{We trained a U-Net convolutional network to predict an image of the average color of the input (top row). Surprisingly, the basic U-Net fails to produce a constant image (c). Similar artifacts appear when using the U-Net for SVBRDF prediction (f). We address this issue by complementing the U-Net with a parallel network that explicitly computes and propagates global features. This approach succeeds in computing the average image (d) and reduces artifacts in SVBRDF maps (g).}
\label{fig:meanExperiment}
\end{figure}

The hourglass shape of the U-Net results in large footprints of the convolution kernels at coarse spatial scales, which in theory provide long-distance dependencies between output pixels. Unfortunately, we found that this multi-scale design is not sufficient to properly fuse information for our problem. We first illustrate this issue on a toy example, where we trained a network to output an image of the average color of the input, as shown in Figure~\ref{fig:meanExperiment} (top row). Surprisingly, the vanilla U-Net performs poorly on this simple task, failing to output a constant-valued image. A similar behavior occurs on our more complex task, where visible residuals of the specular highlight and other fine details pollute the output maps where they should be uniform (Figure~\ref{fig:meanExperiment}, 2nd to 4th row).

In addition, we hypothesize that the ability of the network to compute global information is partly hindered by instance (or batch) normalization, which standardizes the learned features after every convolutional layer by enforcing a mean and standard deviation learned from training data.
In other words, while the normalization is necessary to stabilize training, it actively counters the network's efforts to maintain non-local information about the input image. In fact, instance normalization has been reported to improve artistic style transfer because it eliminates the output's dependence on the input image contrast \cite{Ulyanov_2017_CVPR}. This is the opposite of what we want. \NEW{Unfortunately, while we tried to train a U-Net without normalization, or with a variant of instance normalization without mean subtraction, these networks yielded significant residual shading in all maps.}


We propose a network architecture that simultaneously addresses both of these shortcomings. 
We add a parallel network track alongside the U-Net, which deals with \emph{global} feature vectors instead of 2D feature maps. 
The structure of this global track mirrors that of the main convolutional track, with convolutions changed to fully connected layers and skip connections dropped, and with identical numbers of features. See Figure~\ref{fig:overviewArchi} for an illustration and details of this architecture. The global and convolutional tracks exchange information after every layer as follows:
\begin{itemize} 
	\item Information from the convolutional track flows to the global track via the instance normalization layers. Whereas the standard procedure is to discard the means that are subtracted off the feature maps by instance normalization, we instead incorporate them into the global feature vector using concatenation followed by a fully connected layer and a nonlinearity. For the nonlinearity, we use the Scaled Exponential Linear Unit (SELU) activation function, which is designed to stabilize training for fully connected networks \cite{Klambauer17}.
	\item Information from the global track is injected back into the local track after every convolution, but before the nonlinearity. To do so, we first transform the global features by a fully connected layer, and add them onto each feature map like biases. 
\end{itemize}
\NEW{Our global feature network does not merely preserve the mean signal of a given feature map -- it concatenates the means to form a global feature vector that is processed by fully connected layers before being re-injected in the U-Net at multiple scales.}
Each pair of these information exchanges forms a nonlinear dependency between every pixel, providing the network with means to arrive at a consistent solution by repeatedly transmitting local findings between different regions.
In particular, the common case of near-constant reflectance maps becomes easier for the network to express, as it can source the constant base level from the global features and the fine details from the convolutional maps (Figure~\ref{fig:meanExperiment}g). 


\begin{figure}
\includegraphics[width=\linewidth]{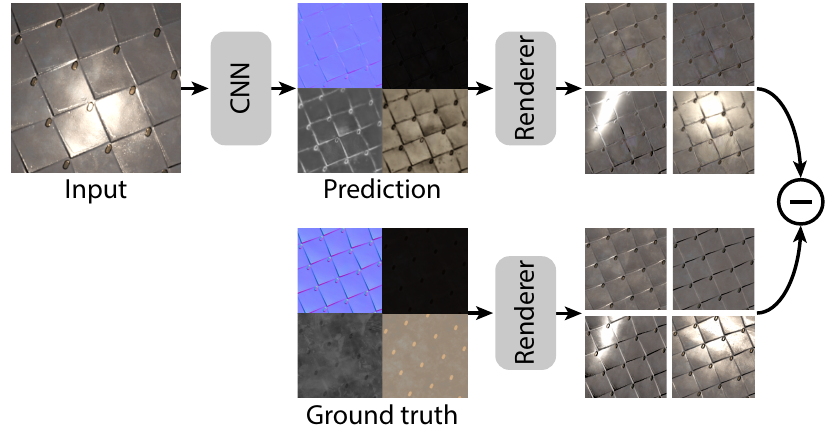}
\caption{Our rendering loss compares the appearance of the predicted SVBRDF and ground truth by rendering both under the same random lighting and viewing configurations.}
\label{fig:renderLoss}
\end{figure}

\begin{figure}


\begin{tabular}{ccccc}

\hspace{-3mm}  \begin{sideways} \hspace{6mm} \small{Normal} \end{sideways} &
\hspace{-3mm}\includegraphics[width=0.3\linewidth]{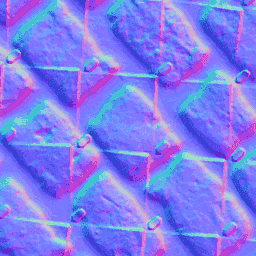} &
\hspace{-3mm}\includegraphics[width=0.3\linewidth]{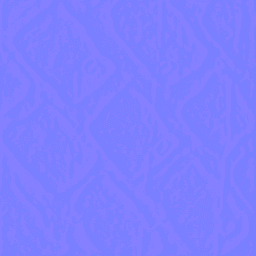} &
\hspace{-3mm}\includegraphics[width=0.3\linewidth]{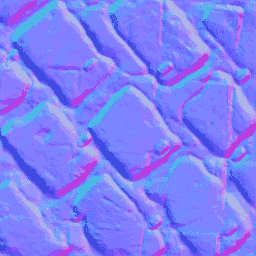} \\

\hspace{-3mm}  \begin{sideways} \hspace{2mm} \small{Diffuse albedo} \end{sideways} &
\hspace{-3mm}\includegraphics[width=0.3\linewidth]{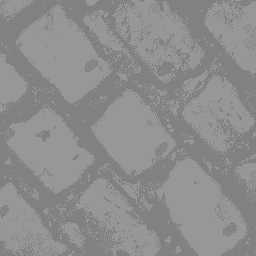} &
\hspace{-3mm}\includegraphics[width=0.3\linewidth]{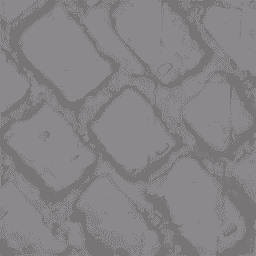} &
\hspace{-3mm}\includegraphics[width=0.3\linewidth]{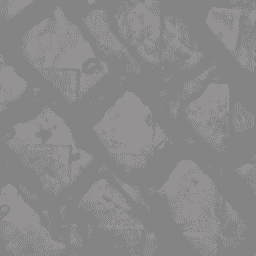} \\

\hspace{-3mm} \begin{sideways} \hspace{5mm} \small{Roughness} \end{sideways} &
\hspace{-3mm}\includegraphics[width=0.3\linewidth]{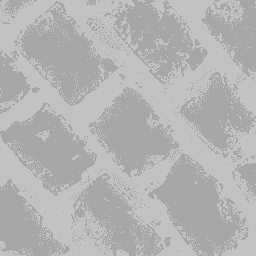} &
\hspace{-3mm}\includegraphics[width=0.3\linewidth]{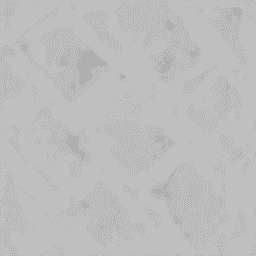} &
\hspace{-3mm}\includegraphics[width=0.3\linewidth]{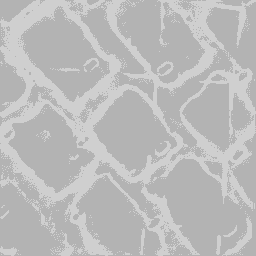} \\

\hspace{-3mm} \begin{sideways} \hspace{2mm} \small{Specular albedo} \end{sideways} &
\hspace{-3mm}\includegraphics[width=0.3\linewidth]{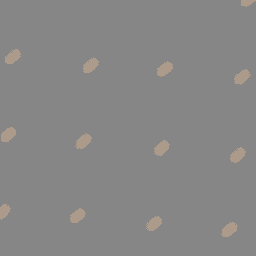} &
\hspace{-3mm}\includegraphics[width=0.3\linewidth]{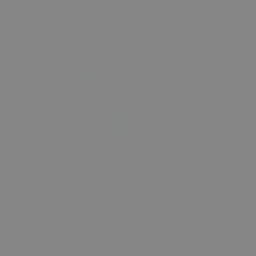} &
\hspace{-3mm}\includegraphics[width=0.3\linewidth]{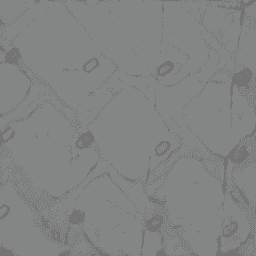} \\ 
\vspace{-3mm}
\\

\hspace{-3mm}  \begin{sideways} \hspace{2mm} \small{Re-rendering} \end{sideways} &
\hspace{-3mm}\includegraphics[width=0.3\linewidth]{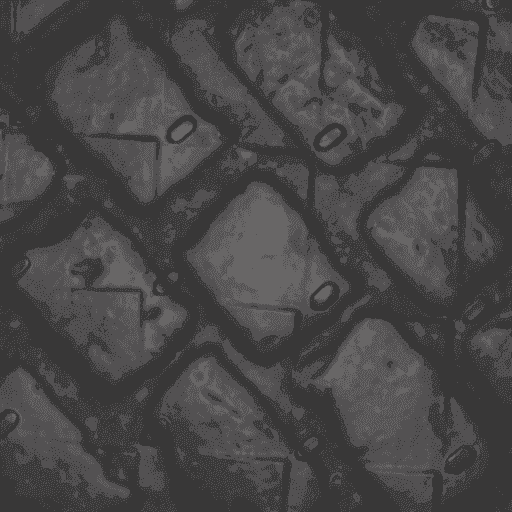} &
\hspace{-3mm}\includegraphics[width=0.3\linewidth]{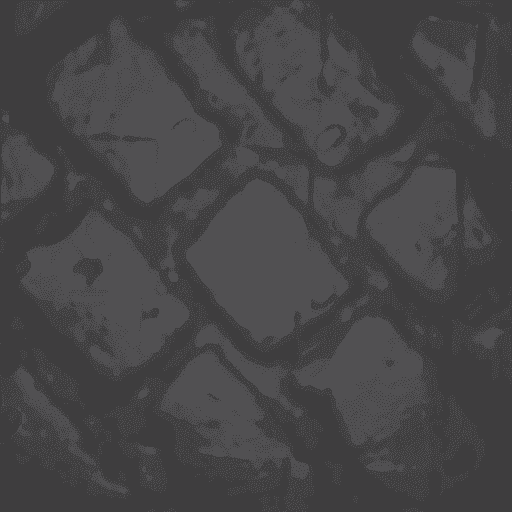} &
\hspace{-3mm}\includegraphics[width=0.3\linewidth]{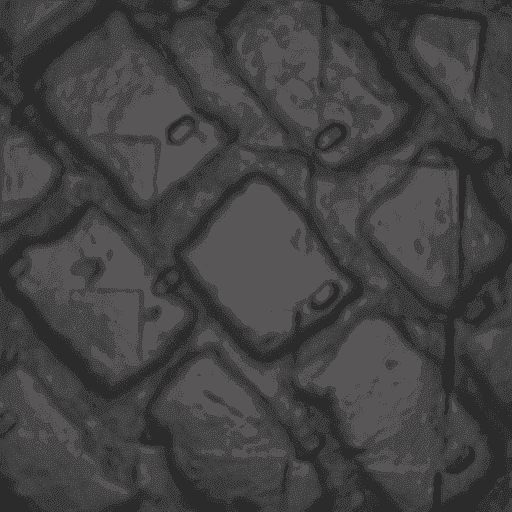} \\

\hspace{-3mm}& 
\hspace{-3mm}\small{(a) GT SVBRDF} & 
\hspace{-3mm}\small{(b) $l_1$ loss} &
\hspace{-3mm}\small{(c) Rendering loss} 

\end{tabular} 

	\caption{When trained with the $l_1$ loss (b), the SVBRDF predicted by the network for a test input image does not accurately reproduce the appearance of the target material when rendered. A network trained using the rendering loss (c) produces an SVBRDF that, while not necessarily identical in terms of the parameter values, reproduces the ground truth rendered appearance well \NEW{(last row)}. 
		}

\label{fig:L1compare}
\end{figure}

\subsection{Rendering Loss}
\label{sec:renderloss}


Our network outputs a set of maps that describe BRDF parameters, such as specular roughness and albedo, at every surface point. The choice of parameterization is arbitrary, as it merely acts as a convenient proxy for the actual object of interest: the spatio-angular appearance of the SVBRDF. In fact, the parameterizations of popular BRDF models arise from a combination of mathematical convenience and relative intuitiveness for artists, and the numerical difference between the parameter values of two (SV)BRDFs is only weakly indicative of their visual similarity.

We propose a loss function that is \emph{independent} of the parameterization of either the predicted or the target SVBRDF, and instead compares their \emph{rendered appearance}. Specifically, any time the loss is evaluated, both the ground truth SVBRDF and the predicted SVBRDF are rendered under identical illumination and viewing conditions, and the resulting images are compared pixel-wise. 
\NEW{We use the same Cook-Torrance BRDF model \shortcite{Cook82} for the ground truth and prediction, but our loss function could equally be used with representations that differ between these two quantities.}


We implement the rendering loss using an in-network renderer, similarly to Aittala et al. \shortcite{Aittala16}. This strategy has the benefits of seamless integration with the neural network training, automatically-computed derivatives, and automatic GPU acceleration. Even complicated shading models are easily expressed in modern deep learning frameworks such as TensorFlow~\cite{tensorflow2015}.
In practice, our renderer acts as a pixel shader that evaluates the rendering equation at each pixel of the SVBRDF, given a pair of view and light directions (Figure~\ref{fig:renderLoss}). Note that this process is performed in the SVBRDF coordinate space, which does not require to output pixels according to the perspective projection of the plane in camera space.

Using a fixed finite set of viewing and lighting directions would make the loss blind to much of the angular space.
Instead, we formulate the loss as the average error over \emph{all} angles, and follow the common strategy of evaluating it stochastically by choosing the angles at random for every training sample, in the spirit of stochastic gradient descent.
To ensure good coverage of typical conditions, we use two sets of lighting and viewing configurations:

\begin{itemize}
	\item The first set of configurations is made of orthographic viewing and lighting directions, sampled independently of one another from the cosine-weighted distribution over the upper hemisphere. The cosine weighting assigns a lower weight to grazing angles, which are observed less often in images due to foreshortening.
	\item While the above configurations cover all angles in theory, in practice it is very unlikely to obtain mirror configurations, which are responsible for visible highlights. Yet, highlights carry rich visual information about material appearance, and should thus contribute to the SVBRDF metric. We ensure the presence of highlights by introducing mirror configurations, where we only sample the lighting direction from the cosine distribution, and use its mirror direction for the viewing direction. We place the origin at a random position on the material plane, and choose independent random distances for both the light and the camera according to the formula $\exp(d)$, where $d \sim \mathrm{Normal} (\mu = 0.5, \sigma = 0.75)$ for a material plane of size $2 \times 2$. The net effect of these configurations is to produce randomly-sized specular highlights at random positions. 
\end{itemize}
We compare the logarithmic values of the renderings using the $l_1$ norm. The logarithm is used to control the potentially extreme dynamic range of specular peaks, and because we are more concerned with relative than absolute errors. To reduce the variance of the stochastic estimate, for every training sample we make 3 renderings in the first configuration and 6 renderings in the second, and average the loss over them. \NEW{We provide a detailed pseudo-code of our rendering loss in supplemental materials.}

Figure~\ref{fig:L1compare} compares the output of our network when trained with a naive $l_1$ loss against the output obtained with our rendering loss. While the $l_1$ loss produces plausible maps when considered in isolation, these maps do not reproduce the appearance of the ground truth once re-rendered. 
In contrast, the rendering loss yields a more faithful reproduction of the ground truth appearance.

\subsection{Training}
We train the network with batch size of 8 for $400{,}000$ iterations, using the Adam optimization algorithm \cite{Kingma14} with a fixed learning rate of $0.00002$. The training takes approximately one week on a TitanX GPU.


\begin{figure}[t]
\center{ 
\begin{tabular}{ccccc}
\hspace{-3mm}\includegraphics[width=0.19\linewidth]{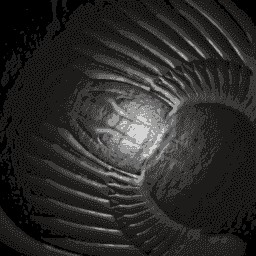} & 
\hspace{-3mm} \includegraphics[width=0.19\linewidth]{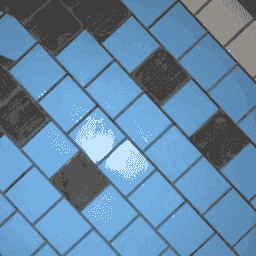} & 
\hspace{-3mm} \includegraphics[width=0.19\linewidth]{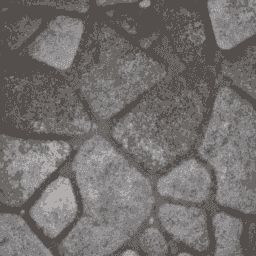} & 
\hspace{-3mm} \includegraphics[width=0.19\linewidth]{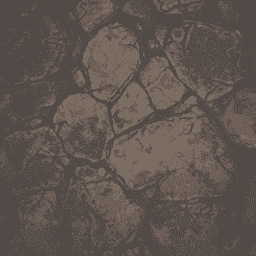} & 
\hspace{-3mm} \includegraphics[width=0.19\linewidth]{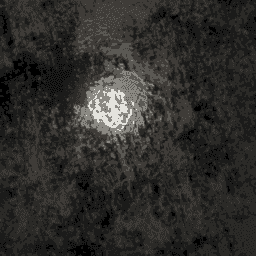} \\
\hspace{-3mm} \small{Leather}&
\hspace{-3mm} \small{Tiles}&
\hspace{-3mm} \small{Stones}&
\hspace{-3mm} \small{Ground}&
\hspace{-3mm} \small{Metal}
 \end{tabular} 
 \begin{tabular}{cccc}
\hspace{-2mm}\includegraphics[width=0.19\linewidth]{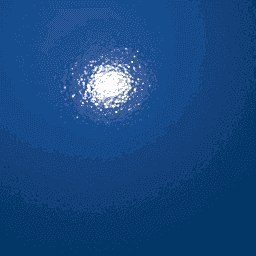} & 
\hspace{-2mm}\includegraphics[width=0.19\linewidth]{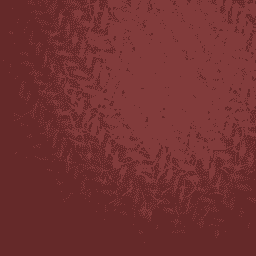} & 
\hspace{-2mm}\includegraphics[width=0.19\linewidth]{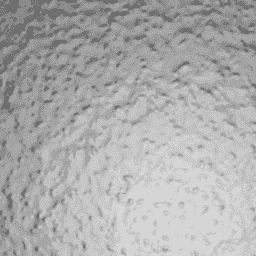} & 
\hspace{-2mm}\includegraphics[width=0.19\linewidth]{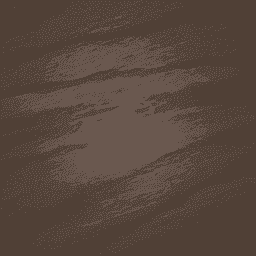} \\
\hspace{-2mm} \small{Plastic}&
\hspace{-2mm} \small{Fabric}&
\hspace{-2mm} \small{Paint}&
\hspace{-2mm} \small{Wood}
 \end{tabular}
}
\caption{Example parametric SVBRDFs for each original material class. We produce our final training set by perturbing and mixing such SVBRDFs.}
\label{fig:SubstanceDiversity}
\end{figure}
\section{Procedural Synthesis of Training Data}
\label{sec:training_data}

\begin{figure}
\begin{tabular}{cccccc}

\hspace{-3mm}  \begin{sideways} \hspace{1mm} \small{Variations} \end{sideways}&
\hspace{-3mm}\includegraphics[width=0.19\linewidth]{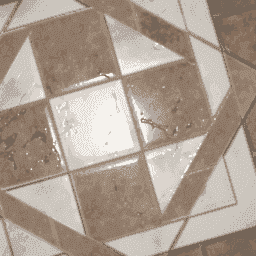} &
\hspace{-3mm}\includegraphics[width=0.19\linewidth]{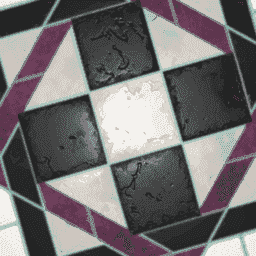} &
\hspace{-3mm}\includegraphics[width=0.19\linewidth]{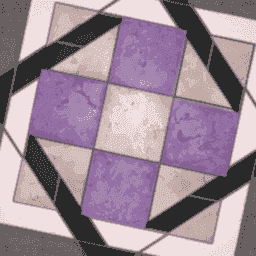} & 
\hspace{-3mm}\includegraphics[width=0.19\linewidth]{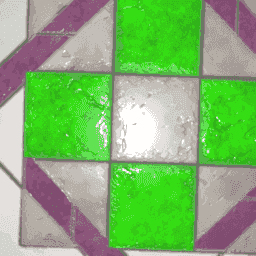} &
\hspace{-3mm}\includegraphics[width=0.19\linewidth]{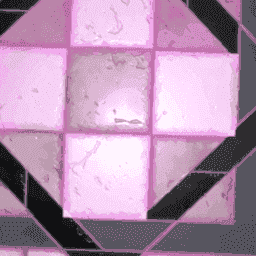} \\

\hspace{-3mm} \begin{sideways} \hspace{1mm} \small{Blendings} \end{sideways}&
\hspace{-3mm}\includegraphics[width=0.19\linewidth]{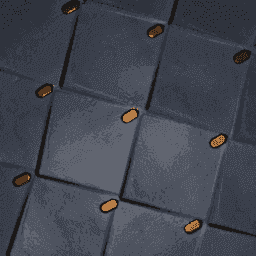} &
\hspace{-3mm}\includegraphics[width=0.19\linewidth]{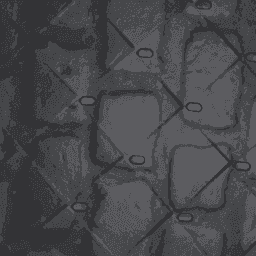} &
\hspace{-3mm}\includegraphics[width=0.19\linewidth]{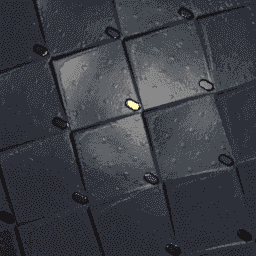} & 
\hspace{-3mm}\includegraphics[width=0.19\linewidth]{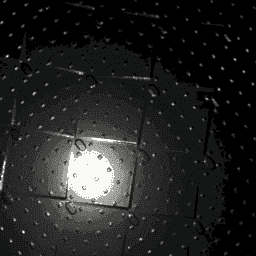} &
\hspace{-3mm}\includegraphics[width=0.19\linewidth]{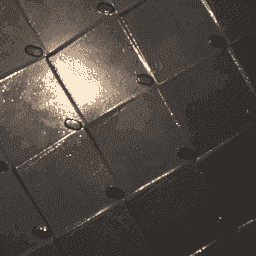} 
 
\end{tabular}
\caption{Data augmentation. We create variations of each parametric SVBRDF by randomly perturbing its parameters (first row).
We additionally augment our dataset by blending pairs of SVBRDFs (second row). Finally, we render each SVBRDF under various orientations, scaling and lighting conditions (both rows).}
\label{fig:dataAugmentation}

\end{figure}

\begin{figure*}
\begin{tabular}{cccccc}

 \includegraphics[width=0.15\linewidth]{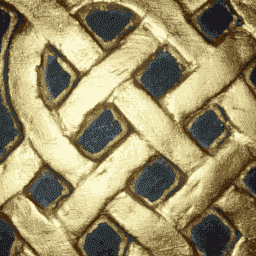} &
\includegraphics[width=0.15\linewidth]{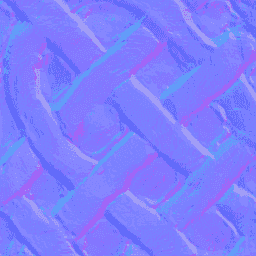} &
\hspace{-3mm}\includegraphics[width=0.15\linewidth]{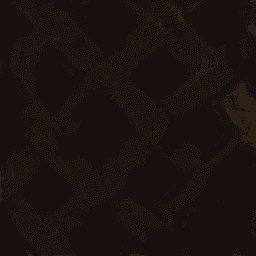} &
\hspace{-3mm}\includegraphics[width=0.15\linewidth]{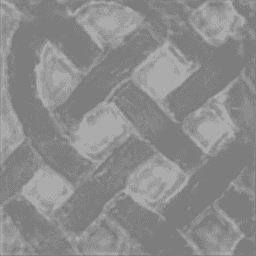} & 
\hspace{-3mm}\includegraphics[width=0.15\linewidth]{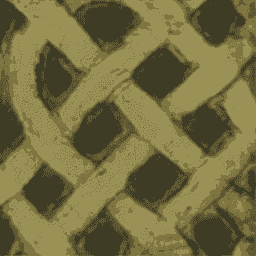} &
\includegraphics[width=0.15\linewidth]{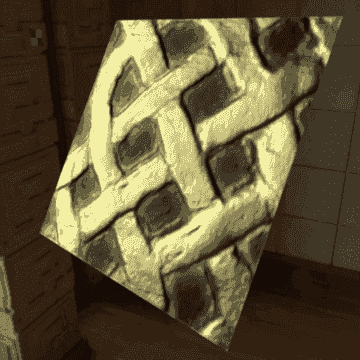} \\

 \includegraphics[width=0.15\linewidth]{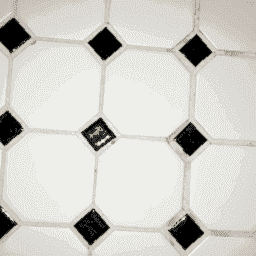} &
\includegraphics[width=0.15\linewidth]{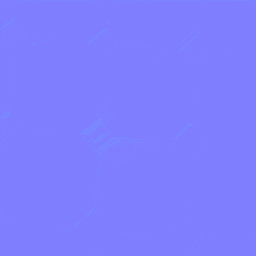} &
\hspace{-3mm}\includegraphics[width=0.15\linewidth]{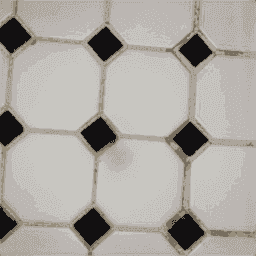} &
\hspace{-3mm}\includegraphics[width=0.15\linewidth]{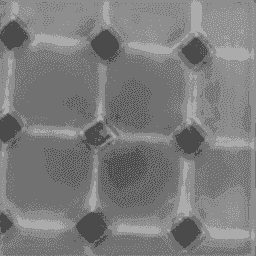} & 
\hspace{-3mm}\includegraphics[width=0.15\linewidth]{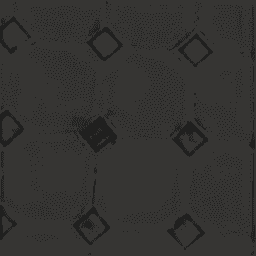} &
\includegraphics[width=0.15\linewidth]{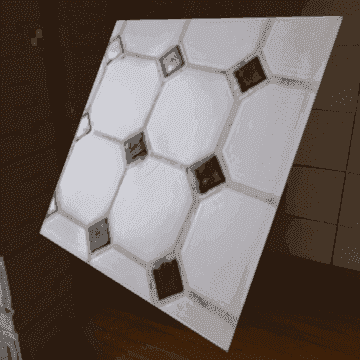} \\
 
 \small{Input} & \small{Normal} & \hspace{-3mm}\small{Diffuse albedo} & \hspace{-3mm}\small{Roughness} & \hspace{-3mm}\small{Specular albedo} & \small{Re-rendering}
\end{tabular}

\vspace{-2mm}
	\caption{Based on the \REM{two}input photographs (left), our method has recovered a set of SVBRDF maps that exhibit strong spatially varying specular roughness and albedo effects. The gold-colored paint (top) and the highly glossy black tiles (bottom) are clearly \REM{distinguished}\NEW{visible} in the re-renderings of SVBRDF under environment illumination (right).}
\label{fig:SpatiallyVaryingSpecular}

\end{figure*}

While several recent papers have shown the potential of synthetic data to train neural networks \cite{Su15,Zhang16,Richter2016}, care must be taken to generate data that is representative of the diversity of real-world materials we want to capture. We address this challenge by leveraging Allegorithmic Substance Share \cite{Substance}, a dataset of more than $800$ procedural SVBRDFs designed by a community of artists from the movie and video game industry. This dataset has several key features relevant to our needs. First, it is representative of the materials artists care about. Second, each SVBRDF is rated by the community, allowing us to select the best ones. Third, each SVBRDF exposes a range of procedural parameters, allowing us to generate variants of them for data augmentation. Finally, each SVBRDF can be converted to the four Cook-Torrance parameter maps we want to predict \cite{Cook82}.


We first curated a set of $155$ high-quality procedural SVBRDFs from $9$ material classes -- paint ($6$), plastic ($5$), leather ($13$), metal ($35$), wood ($23$), fabric ($6$), stone ($25$), ceramic tiles ($29$), ground ($13$), some of which are illustrated in Figure~\ref{fig:SubstanceDiversity}. \NEW{We also selected $12$ challenging procedural SVBRDFs ($6$ metals, $3$ plastics, $3$ woods) to serve as an independent testing set in our comparison to Li et al.~\shortcite{Li17}.}
Together with two artists, we identified the procedural parameters that most influence the appearance of each of \REM{these}\NEW{our training} SVBRDFs. We obtained between $1$ and $36$ parameters per SVBRDF ($7$ on average), for which we manually defined the valid range and default values.

We then performed four types of data augmentation. First, we generated around $1{,}850$ variants of the selected SVBRDFs by applying random perturbations to their important parameters, as illustrated in Figure~\ref{fig:dataAugmentation} (top). 
Second, we generated around $20{,}000$ convex combinations of random pairs of SVBRDFs, which we obtained by $\alpha$-blending their maps.
The mixing greatly increases the diversity of low-level shading effects in the training data, while \REM{straying}\NEW{staying} close to the set of plausible real-world materials, as shown in Figure~\ref{fig:dataAugmentation} (bottom).
Third, we rendered each SVBRDF $10$ times with random lighting, scaling and orientation. Finally, we apply a random crop on each image at training time, so that the network sees slightly different data at each epoch.


The scene we used to render each SVBRDF is composed of a textured plane seen from a fronto-parallel camera and dimensioned to cover the entire image after projection. The light is a small white emitting sphere positioned in a plane parallel to the material sample, at a random offset from the camera center. The camera has a field of view of $50^{\circ}$ to match the typical field of view of cell-phone cameras after cropping to a square, and is positioned at a fixed distance from the material sample. Note that there is a general ambiguity between the scale of the SVBRDF, the distance of the camera, and the strength of the light, which is why we hold the latter parameters fixed. \NEW{However, since such parameters are unknown in our casual capture scenario, the albedo maps we obtain from real pictures at test time are subject to an arbitrary, global scale factor.}

We used the Mitsuba renderer \cite{Mitsuba}, for which we implemented the Cook-Torrance BRDF model \shortcite{Cook82} with GGX normal distribution \cite{Walter07} to match the model used in Allegorithmic Substance. We rendered each SVBRDF as a linear low-dynamic range image, similar to gamma-inverted photographs captured with a cell-phone.
We also used Mitsuba to render the parameter maps after random scaling and rotation of the material sample, which ensures that the maps are aligned with the material rendering and that the normal map is expressed in screen coordinate space rather than texture coordinate space.
Our entire dataset of around $200{,}000$ SVBRDFs took around $16$ hours to generate on a cluster of 40 CPUs.

\begin{figure*}[t]
\begin{tabular}{cccccc}
 \includegraphics[align=c, width=0.15\textwidth]{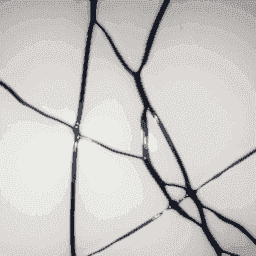} &
 \hspace{-5mm} \includegraphics[align=c, width=0.15\textwidth]{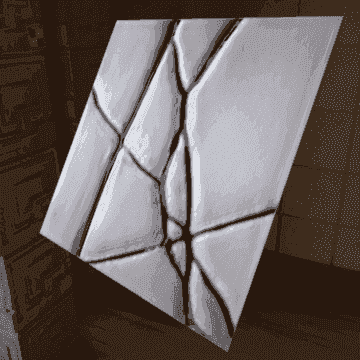} &
  \includegraphics[align=c, width=0.15\textwidth]{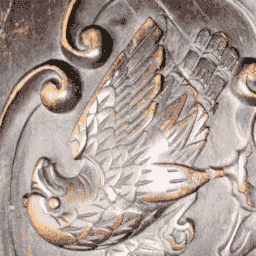} &
 \hspace{-5mm} \includegraphics[align=c, width=0.15\textwidth]{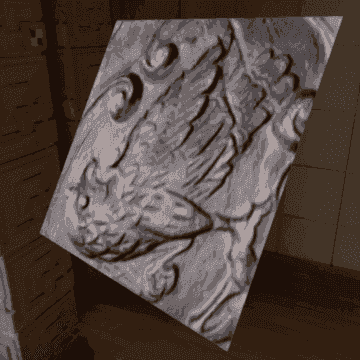} &
  \includegraphics[align=c, width=0.15\textwidth]{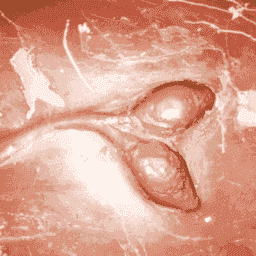} &
 \hspace{-5mm} \includegraphics[align=c, width=0.15\textwidth]{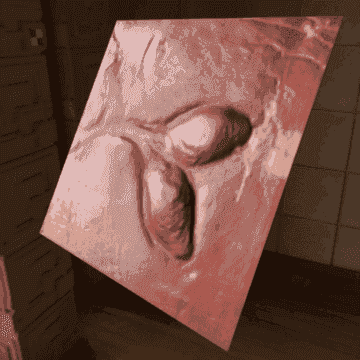} \\ \\
 \includegraphics[align=c, width=0.15\textwidth]{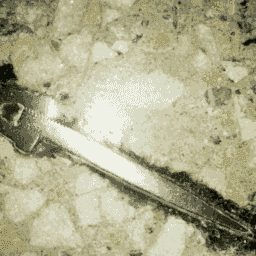} &
 \hspace{-5mm} \includegraphics[align=c, width=0.15\textwidth]{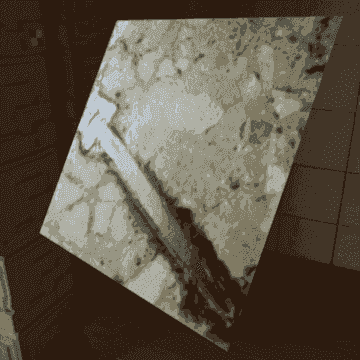} &
  \includegraphics[align=c, width=0.15\textwidth]{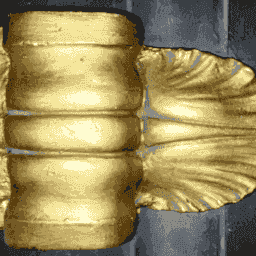} &
 \hspace{-5mm} \includegraphics[align=c, width=0.15\textwidth]{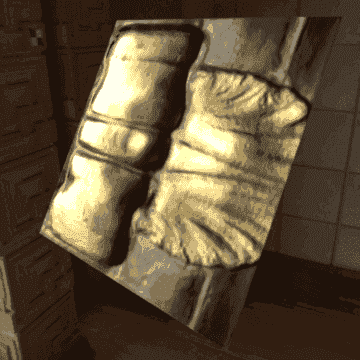} &
  \includegraphics[align=c, width=0.15\textwidth]{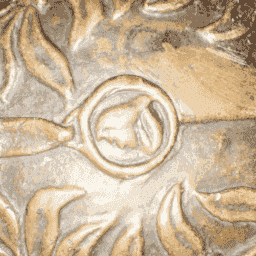} &
 \hspace{-5mm} \includegraphics[align=c, width=0.15\textwidth]{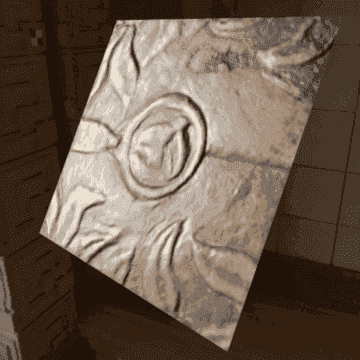} \\ \\
 \includegraphics[align=c, width=0.15\textwidth]{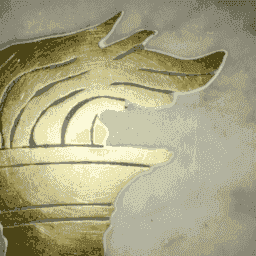} &
 \hspace{-5mm} \includegraphics[align=c, width=0.15\textwidth]{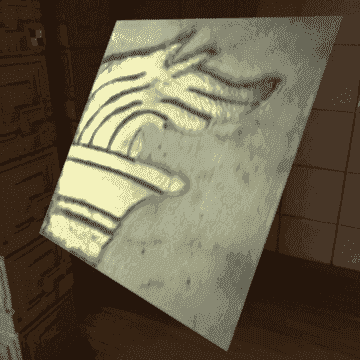} &
  \includegraphics[align=c, width=0.15\textwidth]{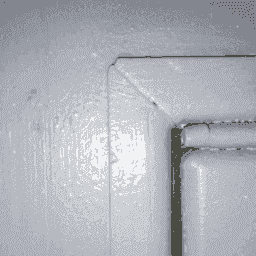} &
 \hspace{-5mm} \includegraphics[align=c, width=0.15\textwidth]{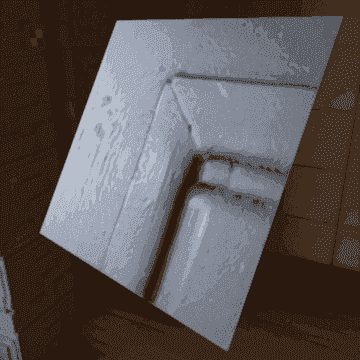} &
  \includegraphics[align=c, width=0.15\textwidth]{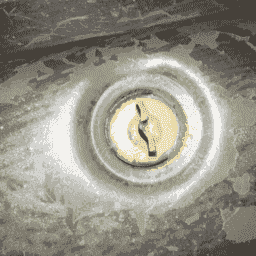} &
 \hspace{-5mm} \includegraphics[align=c, width=0.15\textwidth]{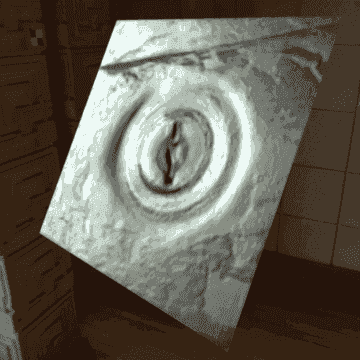} \\ \\
\end{tabular}

\vspace{-4mm}
	\caption{A selection of results from our method on real-world photographs. In each image pair, the left image is a photograph of a surface, and the right image is a re-rendering of the SVBRDF \NEW{inferred} from that image. The illumination environment in the re-renderings is an interior space with a large window on the left. See supplemental materials for additional results and animated re-renderings.}

\label{fig:ResultMatrix}
\end{figure*}

\begin{figure*}

\begin{tabular}{cc}

 \begin{tabular}{c}
\includegraphics[align=c, width=0.15\linewidth]{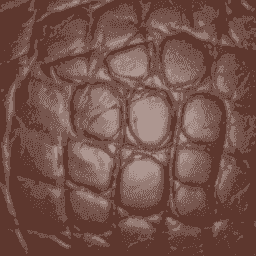} \\
\small{Input (Leather)}
\end{tabular}

&

\begin{tabular}{ccccc}

\hspace{-3mm} \begin{sideways} \hspace{10mm} \small{BTF} \end{sideways}&
\hspace{-3mm}\includegraphics[width=0.15\linewidth]{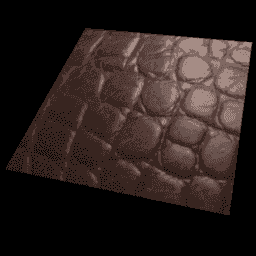} &
\hspace{-3mm}\includegraphics[width=0.15\linewidth]{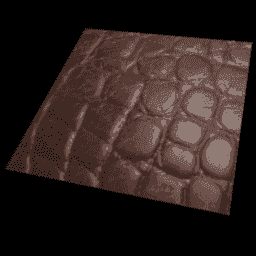} & 
\hspace{-3mm}\includegraphics[width=0.15\linewidth]{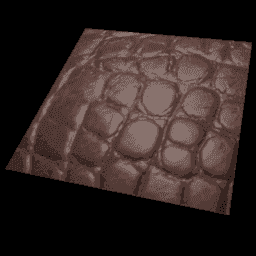} &
\hspace{-3mm}\includegraphics[width=0.15\linewidth]{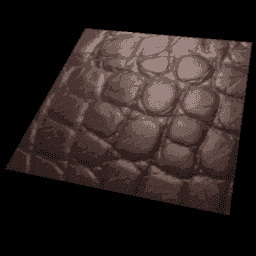} \\

\hspace{-3mm} \begin{sideways} \hspace{10mm} \small{Ours} \end{sideways}&
\hspace{-3mm}\includegraphics[width=0.15\linewidth]{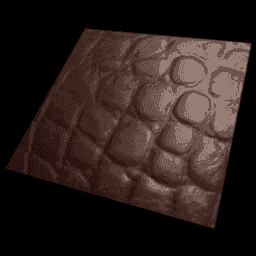} &
\hspace{-3mm}\includegraphics[width=0.15\linewidth]{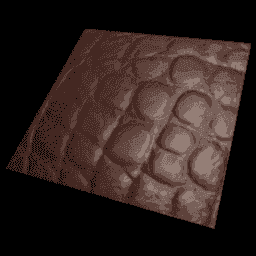} & 
\hspace{-3mm}\includegraphics[width=0.15\linewidth]{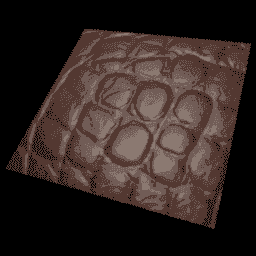} &
\hspace{-3mm}\includegraphics[width=0.15\linewidth]{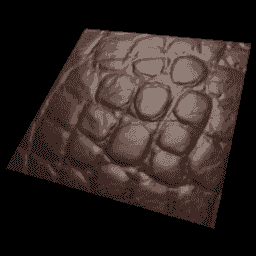} 

\end{tabular}
\\
 \begin{tabular}{c}
\includegraphics[align=c, width=0.15\linewidth]{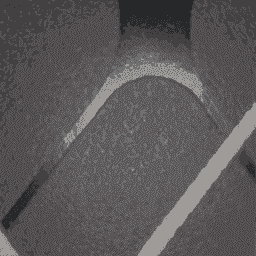} \\
\small{Input (Wallpaper)}
\end{tabular}

&

\begin{tabular}{ccccc}

\hspace{-3mm} \begin{sideways} \hspace{10mm} \small{BTF} \end{sideways}&
\hspace{-3mm}\includegraphics[width=0.15\linewidth]{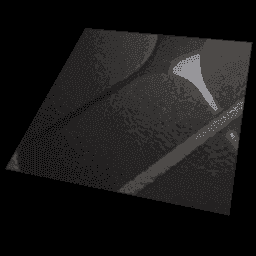} &
\hspace{-3mm}\includegraphics[width=0.15\linewidth]{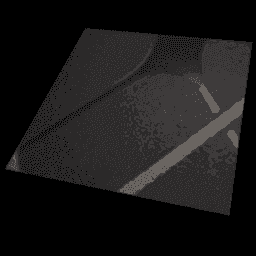} & 
\hspace{-3mm}\includegraphics[width=0.15\linewidth]{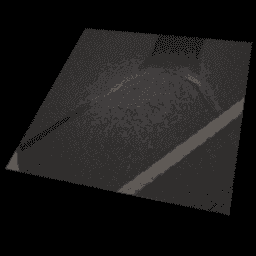} &
\hspace{-3mm}\includegraphics[width=0.15\linewidth]{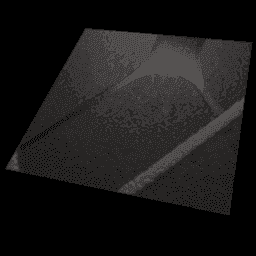} \\

\hspace{-3mm} \begin{sideways} \hspace{10mm} \small{Ours} \end{sideways}&
\hspace{-3mm}\includegraphics[width=0.15\linewidth]{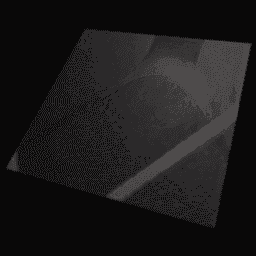} &
\hspace{-3mm}\includegraphics[width=0.15\linewidth]{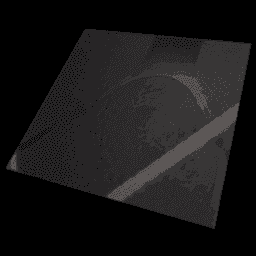} & 
\hspace{-3mm}\includegraphics[width=0.15\linewidth]{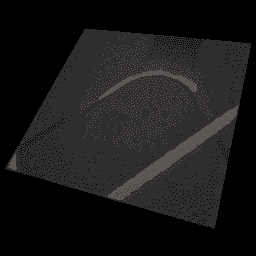} &
\hspace{-3mm}\includegraphics[width=0.15\linewidth]{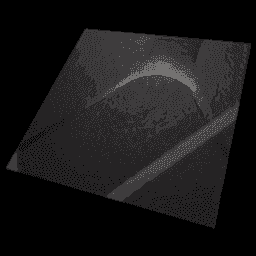} 

\end{tabular}

\end{tabular} 

\vspace{-1mm}
\caption{\NEW{Comparison between relighting of our predictions and of measured BTFs \protect\cite{weinmann2014}.}}
\label{fig:RelightingExperimentBTF}
\end{figure*}

\begin{figure*}

\begin{tabular}{cc}

\begin{tabular}{c}
\includegraphics[align=c, width=0.15\linewidth]{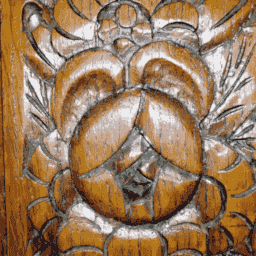} \\
\small{Input (Wood)}
\end{tabular}

&

\begin{tabular}{ccccc}

\hspace{-3mm} \begin{sideways} \hspace{5mm} \small{Real pictures} \end{sideways}&
\hspace{-3mm}\includegraphics[width=0.15\linewidth]{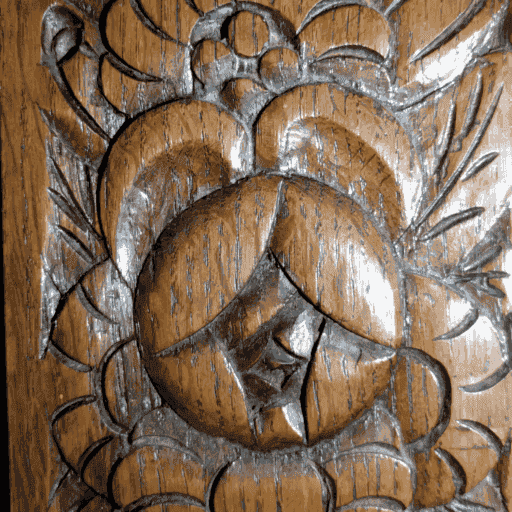} &
\hspace{-3mm}\includegraphics[width=0.15\linewidth]{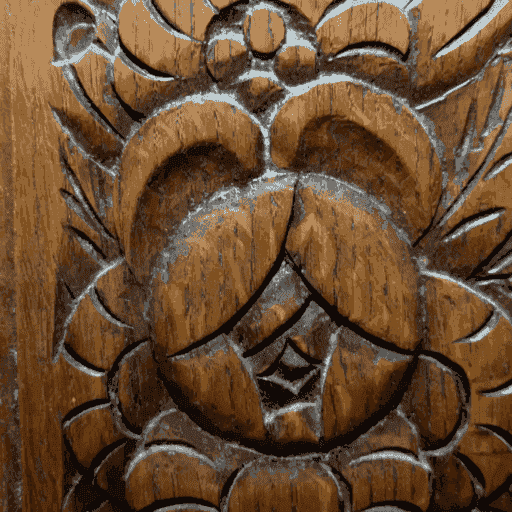} & 
\hspace{-3mm}\includegraphics[width=0.15\linewidth]{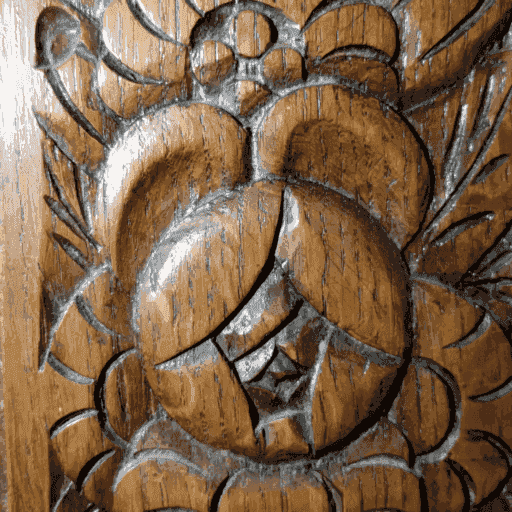} &
\hspace{-3mm}\includegraphics[width=0.15\linewidth]{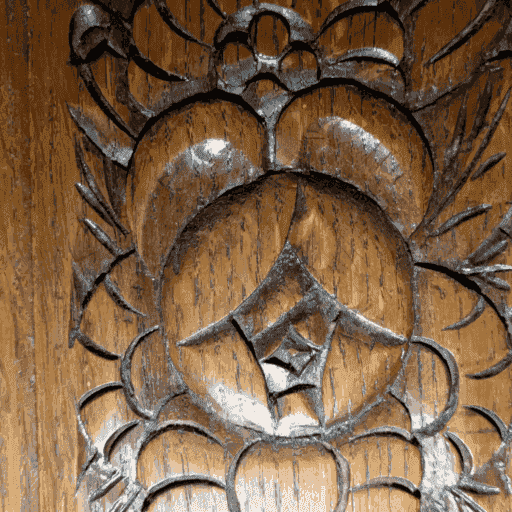} \\

\hspace{-3mm} \begin{sideways} \hspace{10mm} \small{Ours} \end{sideways}&
\hspace{-3mm}\includegraphics[width=0.15\linewidth]{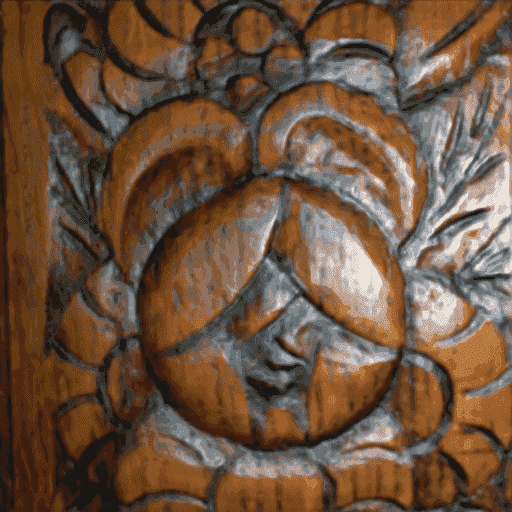} &
\hspace{-3mm}\includegraphics[width=0.15\linewidth]{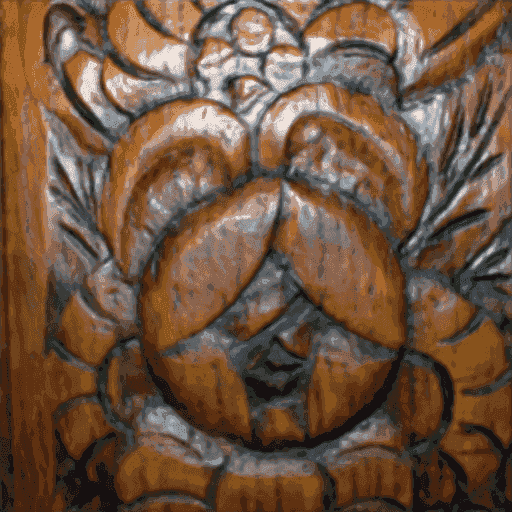} & 
\hspace{-3mm}\includegraphics[width=0.15\linewidth]{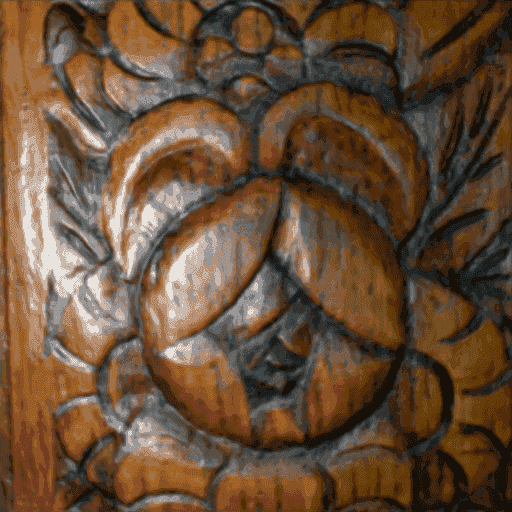} &
\hspace{-3mm}\includegraphics[width=0.15\linewidth]{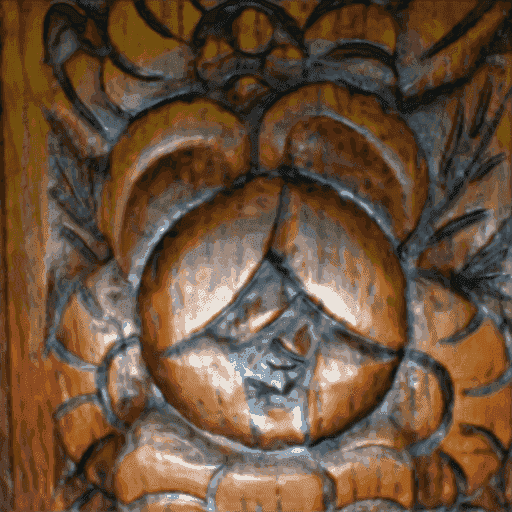} 

\end{tabular}

\end{tabular} 

\vspace{-1mm}
\caption{Comparison between relighting of our prediction and real pictures under approximately the same lighting configurations. We adjusted the white balance of the results to best match the one of the input.}
\label{fig:RelightingExperimentPhoto}
\end{figure*}

\section{Evaluation}

We now evaluate our approach on real-world photographs and compare it with recent methods for single-image SVBRDF capture.
We refer the reader to the supplemental material for an extensive set of results for hundreds of materials, including all estimated SVBRDF maps and further re-renderings. 
In particular, animations with moving light sources demonstrate that the solutions work equally well in a variety of lighting conditions.
The supplemental material also includes additional comparisons with previous work.




\subsection{Real-world photographs}

We used regular cell phones (iPhone SE and Nexus 5X) and their built-in flash units to capture a dataset of nearly 350 materials on which we applied our method. We cropped the images to approximate the field of view used in the training data. The dataset includes samples from a large variety of materials found in domestic, office and public interiors, as well as outdoors. In fact, most of the photographs were shot during a casual walk-around within the space of a few hours. 

Figures \ref{fig:teaser} and \ref{fig:ResultMatrix} show a selection of representative pairs of input photographs, and corresponding re-renderings of the results under novel environment illumination. 
The results demonstrate that the method successfully reproduces a rich set of reflectance effects for metals, plastics, paint, wood and various more exotic substances, often mixed together in the same image. We found it to perform particularly well on materials exhibiting bold large-scale features, where the normal maps capture sharp and complex geometric shapes from the photographed surfaces.

Figure \ref{fig:SpatiallyVaryingSpecular} shows our result for two materials with interesting spatially varying specularity behavior. The method has successfully identified the gold paint in the specular albedo map, and the different roughness levels of the black and white tiles. The latter feature shows good consistency across the spatially distant black squares, and we find it particularly impressive that the low roughness level was apparently resolved based on the small highlight cues on the center tile and the edges of the outer tiles.  For most materials, the specular albedo is resolved as monochrome, as it should be. Similar globally consistent behavior can be seen across the result set: cues from sparsely observed specular highlights often inform the specularity across the entire material. 

Note that our dataset contains several duplicates, i.e. multiple shots of the same material taken from slightly different positions. Their respective SVBRDF solutions generally show good consistency among each other. 
\NEW{We also captured a few pictures with an SLR camera, for which the flash is located further away from the lens than cell phones. We provide the resulting predicted maps in supplemental materials, showing that our method is robust to varying positions of the flash.}

\subsection{Comparisons}
\label{sec:comparisons}

\subsubsection{Relighting}
\NEW{Figure~\ref{fig:RelightingExperimentBTF} provides a qualitative
comparison between renderings of our predictions and renderings of 
measured Bidirectional Texture Functions (BTFs) \cite{weinmann2014} under the same lighting conditions. 
While BTFs are not parameterized according to the 4 maps we estimate, they capture ground-truth appearance from arbitrary view and lighting conditions, which ultimately is the quantity we wish to reproduce. Our method provides a faithful reproduction of the appearance of the leather. It also captures well the spatially-varying specularity of the wallpaper, even though it produces slightly more blurry highlights. Please refer to supplemental materials for additional results on $20$ BTFs.}

In addition, Figure~\ref{fig:RelightingExperimentPhoto} compares renderings of our predictions 
with real photographs under approximately similar lighting conditions. \NEW{Our method is especially effective at capturing the normal variations of this wood carving.}

\begin{figure*}[t]

\begin{tabular}{cc}

 \begin{tabular}{c}
\includegraphics[align=c, width=0.15\linewidth]{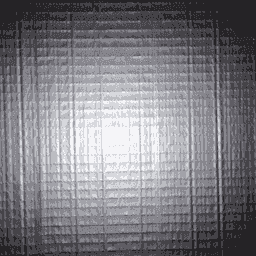} \\
\small{Input (Tape)}
\end{tabular}

 \begin{tabular}{cccccc}
\begin{sideways} \hspace{-11mm} \small{\cite{Aittala16}} \end{sideways} &
\hspace{-3mm}  \includegraphics[align=c, width=0.15\linewidth]{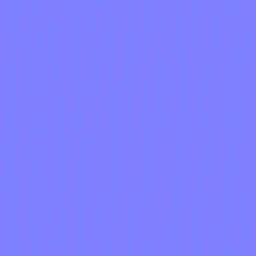} & 
\hspace{-3mm} \includegraphics[align=c, width=0.15\linewidth]{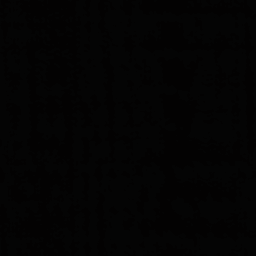} & 
\hspace{-3mm} \includegraphics[align=c, width=0.15\linewidth]{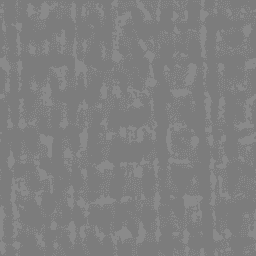} & 
\hspace{-3mm} \includegraphics[align=c, width=0.15\linewidth]{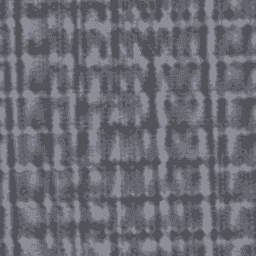} & 
 \includegraphics[align=c, width=0.15\linewidth]{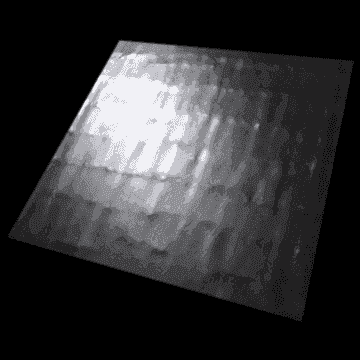}\\ 
\vspace{-3mm}
 \\
 
 \begin{sideways} \hspace{-11mm} \small{\cite{Aittala15}} \end{sideways} &
\hspace{-3mm}  \includegraphics[align=c, width=0.15\linewidth]{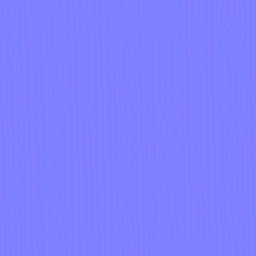} & 
\hspace{-3mm} \includegraphics[align=c, width=0.15\linewidth]{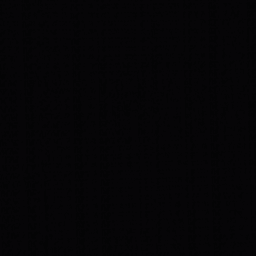} & 
\hspace{-3mm} \includegraphics[align=c, width=0.15\linewidth]{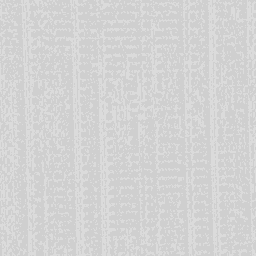} & 
\hspace{-3mm} \includegraphics[align=c, width=0.15\linewidth]{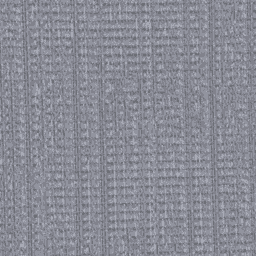} & 
 \includegraphics[align=c, width=0.15\linewidth]{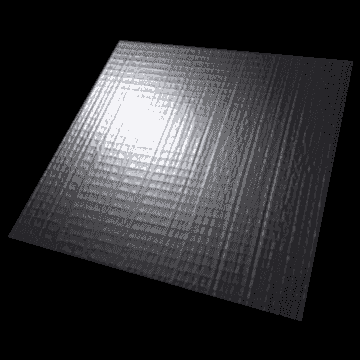}\\ 
\vspace{-3mm}
 \\

\begin{sideways} \hspace{-8mm}  \small{Our prediction} \end{sideways} &
\hspace{-3mm}  \includegraphics[align=c, width=0.15\linewidth]{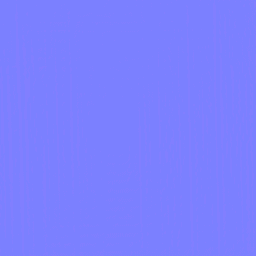} & 
\hspace{-3mm} \includegraphics[align=c, width=0.15\linewidth]{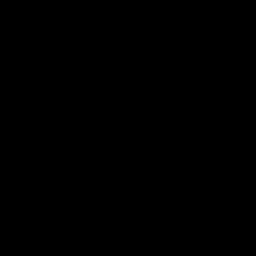} & 
\hspace{-3mm} \includegraphics[align=c, width=0.15\linewidth]{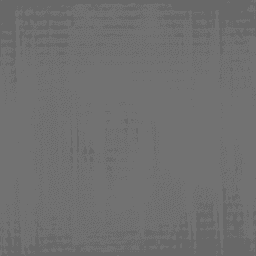} & 
\hspace{-3mm} \includegraphics[align=c, width=0.15\linewidth]{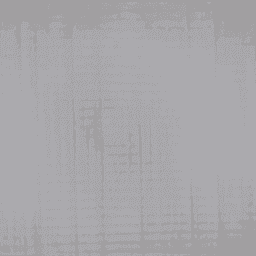} & 
 \includegraphics[align=c, width=0.15\linewidth]{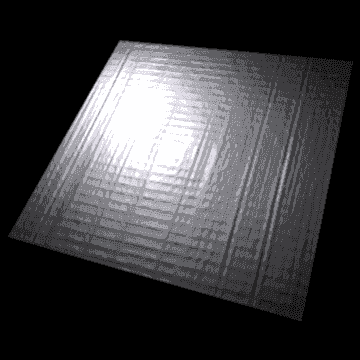}\\ 
 & \hspace{-3mm} \small{Normal} &\hspace{-3mm}  \small{Diffuse albedo} &\hspace{-3mm}  \small{Roughness} & \hspace{-3mm} \small{Specular albedo} &  \small{Re-rendering} 

 \end{tabular}\\

\end{tabular}

\vspace{-2mm}
\caption{Comparison with Aittala et al. \protect \shortcite{Aittala15,Aittala16}. Note that the maps (other than the normals) are not directly comparable due to different parametrization of the BRDF models. \REM{Our solution also represents the entire input image, whereas}The solution of Aittala et al. \protect \shortcite{Aittala16} corresponds to a small region of about $15 \%$ of the image dimension, intended to be repeated by texture synthesis or tiling. 
\NEW{The earlier method by Aittala et al. \protect \shortcite{Aittala15} captures the entire input image but requires an additional no-flash picture for guidance.}
In contrast, our method reproduces the large-scale features well, and is applicable to non-repetitive materials \NEW{captured with a single flash picture.}}

\label{fig:ComparisonMiika}
\end{figure*}

\subsubsection{Aittala et al. \shortcite{Aittala15,Aittala16}}
The method by Aittala et al.~\shortcite{Aittala16} is the most related to ours in terms of input, since it also computes an SVBRDF representation from a single flash-lit photograph.
However, Aittala et al.~\shortcite{Aittala16} exploit redundancy in the input picture by assuming that the material is \emph{stationary}, \emph{i.e.} consists of small textural features that repeat throughout the image. 

We compare our method to theirs by feeding photographs from their dataset to our network (Figure~\ref{fig:ComparisonMiika} \NEW{and supplemental materials}). Despite the similar input, the two approaches produce different outputs: whereas we produce a map that represents the entire input photo downsampled to $256 \times 256$, their method produces a tile that represents a small piece of the texture at high resolution. Furthermore, the BRDF models used by the methods are different. To aid comparison, we show re-renderings of the material predicted by each method under identical novel lighting conditions.

\REM{For \emph{Tape}, }Both methods produce a good result, but show a clearly different character. The method of Aittala et al.~\shortcite{Aittala16} recovers sharp textural details that are by construction similar across the image. For the same reason, their solution cannot express larger-scale variations, and the result is somewhat repetitive. In contrast, our solution shows more interesting large-scale variations across the image, but lacks some detail and consistency in the local features. 
\REM{\emph{Leather} contains a significant amount of larger-scale non-stationary detail,} 

\REM{which the method of Aittala et al.~\shortcite{Aittala16} cannot capture faithfully.}

\REM{This leads to an overly monotonous result when the map is tiled }

\REM{or synthesized to cover a larger surface region. Our method does not }

\REM{suffer from this limitation, and produces an overall more natural }

\REM{reproduction of the leather.}

Most of our real-world test images violate the stationarity requirement, and as such would not be suitable for the method of Aittala et al.~\shortcite{Aittala16}. Our method also has the advantage in speed: whereas Aittala et al.~\shortcite{Aittala16} use an iterative optimization that takes more than an hour per material sample, our feedforward network evaluation is practically instant.

\NEW{Figure~\ref{fig:ComparisonMiika} also contains results obtained with an earlier method by Aittala et al.~\shortcite{Aittala15}. This method also assumes stationary materials, and requires an additional no-flash picture to identify repetitive details and their large-scale variations. Our approach produces similar results from a single image, although at a lower resolution.}

\subsubsection{Li et al.~\shortcite{Li17}}
\begin{figure*}[t]
 \begin{tabular}{ccccccc}

\hspace{-9mm}\begin{sideways} \hspace{-8mm}  \small{\cite{Li17}} \end{sideways} &
\hspace{-6mm} \includegraphics[align=c, width=0.14\linewidth]{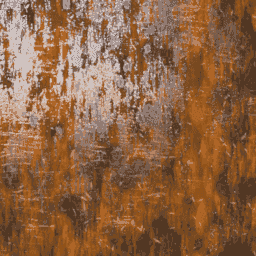} & 
 \includegraphics[align=c, width=0.14\linewidth]{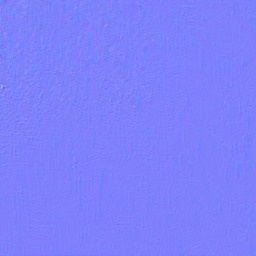} & 
\hspace{-3mm} \includegraphics[align=c, width=0.14\linewidth]{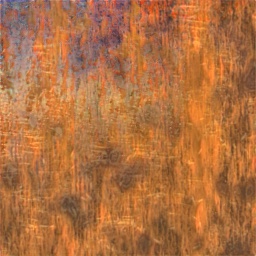} & 
\hspace{-3mm} \includegraphics[align=c, width=0.14\linewidth]{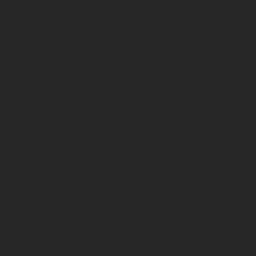} & 
\hspace{-3mm} \includegraphics[align=c, width=0.14\linewidth]{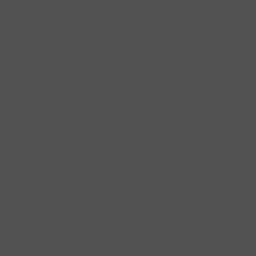} & 
 \includegraphics[align=c, width=0.14\linewidth]{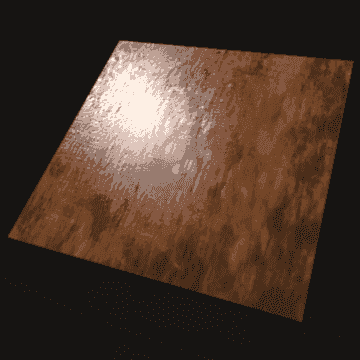} \\
\vspace{-3mm}

\\
\hspace{-9mm}\begin{sideways} \hspace{-7mm} \small{Ground truth} \end{sideways} & & 
 \includegraphics[align=c, width=0.14\linewidth]{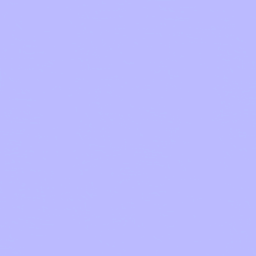} & 
\hspace{-3mm} \includegraphics[align=c, width=0.14\linewidth]{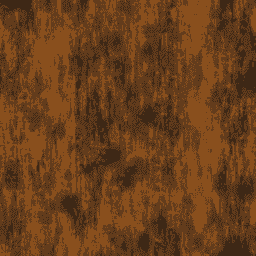} & 
\hspace{-3mm} \includegraphics[align=c, width=0.14\linewidth]{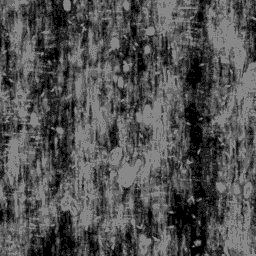} & 
\hspace{-3mm} \includegraphics[align=c, width=0.14\linewidth]{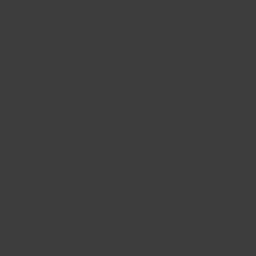} & 
 \includegraphics[align=c, width=0.14\linewidth]{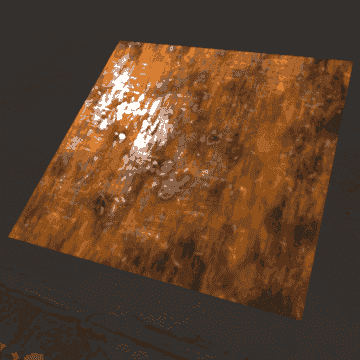}\\ 
\vspace{-3mm}
 
\\
\hspace{-9mm}\begin{sideways} \hspace{-8mm}  \small{Our approach} \end{sideways} &
\hspace{-6mm} \includegraphics[align=c, width=0.14\linewidth]{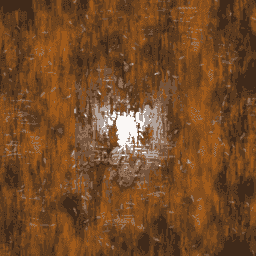} & 
 \includegraphics[align=c, width=0.14\linewidth]{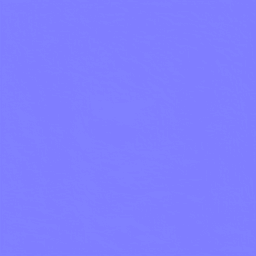} & 
\hspace{-3mm} \includegraphics[align=c, width=0.14\linewidth]{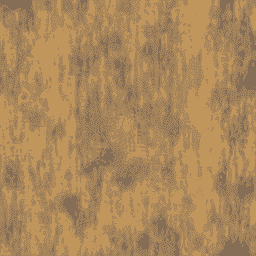} & 
\hspace{-3mm} \includegraphics[align=c, width=0.14\linewidth]{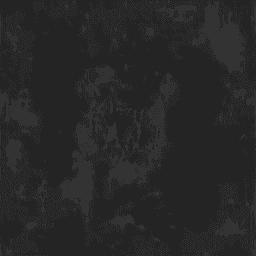} & 
\hspace{-3mm} \includegraphics[align=c, width=0.14\linewidth]{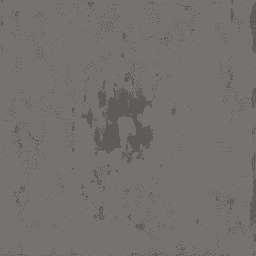} & 
 \includegraphics[align=c, width=0.14\linewidth]{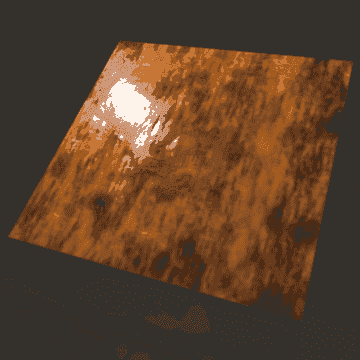} 
\\
\vspace{-2mm}
\\

\hspace{-9mm}\begin{sideways} \hspace{-8mm}  \small{\cite{Li17}} \end{sideways} &
\hspace{-6mm} \includegraphics[align=c, width=0.14\linewidth]{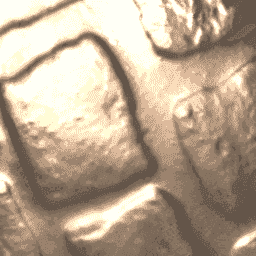} & 
 \includegraphics[align=c, width=0.14\linewidth]{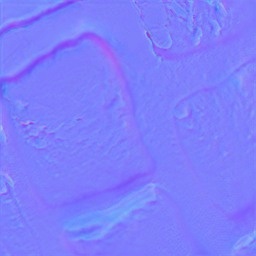} & 
\hspace{-3mm} \includegraphics[align=c, width=0.14\linewidth]{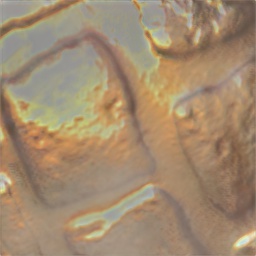} & 
\hspace{-3mm} \includegraphics[align=c, width=0.14\linewidth]{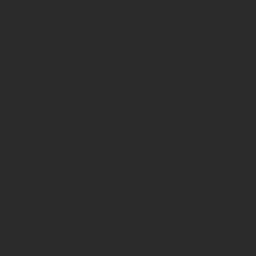} & 
\hspace{-3mm} \includegraphics[align=c, width=0.14\linewidth]{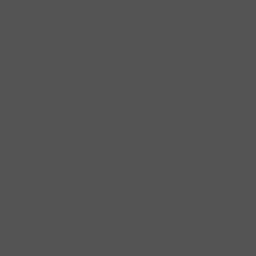} & 
 \includegraphics[align=c, width=0.14\linewidth]{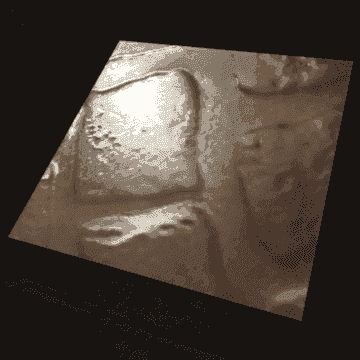} \\
\vspace{-3mm}

\\
\hspace{-9mm}\begin{sideways} \hspace{-7mm} \small{Ground truth} \end{sideways} & & 
 \includegraphics[align=c, width=0.14\linewidth]{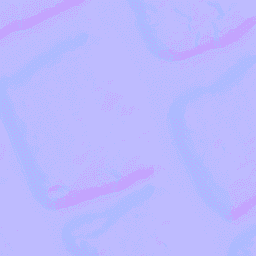} & 
\hspace{-3mm} \includegraphics[align=c, width=0.14\linewidth]{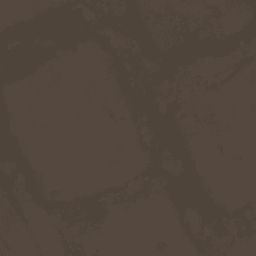} & 
\hspace{-3mm} \includegraphics[align=c, width=0.14\linewidth]{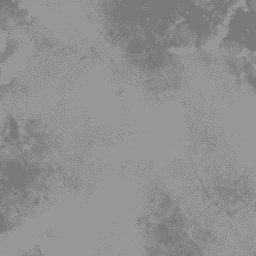} & 
\hspace{-3mm} \includegraphics[align=c, width=0.14\linewidth]{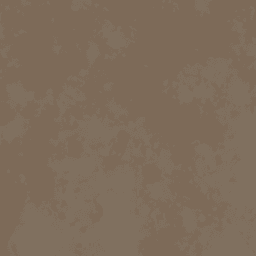} & 
 \includegraphics[align=c, width=0.14\linewidth]{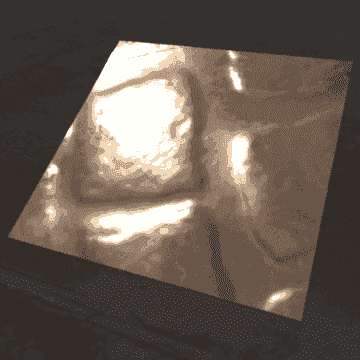}\\ 
\vspace{-3mm}
 
\\
\hspace{-9mm}\begin{sideways} \hspace{-8mm}  \small{Our approach} \end{sideways} &
\hspace{-6mm} \includegraphics[align=c, width=0.14\linewidth]{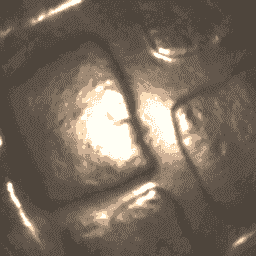} & 
 \includegraphics[align=c, width=0.14\linewidth]{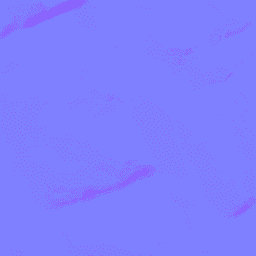} & 
\hspace{-3mm} \includegraphics[align=c, width=0.14\linewidth]{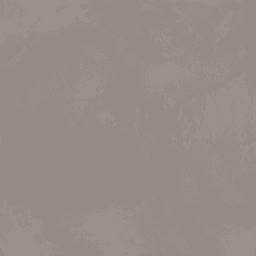} & 
\hspace{-3mm} \includegraphics[align=c, width=0.14\linewidth]{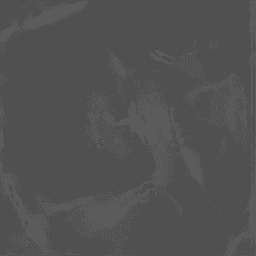} & 
\hspace{-3mm} \includegraphics[align=c, width=0.14\linewidth]{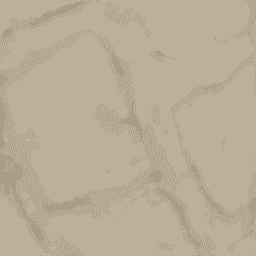} & 
\includegraphics[align=c, width=0.14\linewidth]{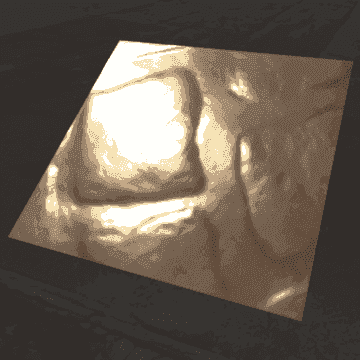} 
\\
&\hspace{-6mm}\small{Input}&\small{Normal}&\hspace{-3mm}\small{Diffuse albedo}&\hspace{-3mm}\small{Roughness}&\hspace{-3mm}\small{Specular albedo}&\small{Re-rendering}
\\

 \end{tabular}

\vspace{-2mm}
\caption{Comparison with Li et al. \protect \shortcite{Li17} on synthetic data. As the methods produce output data using different BRDF models, the values of the maps of Li et al. \protect \shortcite{Li17} should not be compared directly to ours or the ground truth. We show them to aid qualitative evaluation of the the spatial variation. To facilitate comparison, we rendered the ground truth and each result under novel illumination conditions (right). The renderings for the results of Li et al. \protect \shortcite{Li17} were made with a lower exposure due to different albedo magnitudes predicted by the methods. The input images (left) were rendered under flash lighting for our method and under environment lighting for the method by Li et al., in agreement with the type of input assumed by each method.}

\label{fig:SyntheticComparisonMicrosoft}
\end{figure*}

The method by Li et al.~\shortcite{Li17} is based on a similar U-Net convolutional network as ours. However, it has been designed to process pictures captured under environment lighting rather than flash lighting, and it predicts a constant specular albedo and roughness instead of spatially-varying maps. We first compare the two methods on \REM{synthetic images}\NEW{our synthetic test set} for which we have the ground truth SVBRDFs (Figure~\ref{fig:SyntheticComparisonMicrosoft} \NEW{and Table~\ref{table:quantitativeComparison}}). For a fair comparison, we tested the method by Li et al. on several renderings of the ground truth, using different environment maps and different orientations. We then selected the input image that gave the best outcome. We compare the results of the two methods \NEW{qualitatively} with re-renderings under a mixed illumination composed of an environment map enriched with a flash light, so as to ensure that neither method has an advantage. \NEW{For quantitative comparison, we compute the RMSE of each individual map, as well as the RMSE of re-renderings averaged over multiple point lighting conditions; our results have systematically lower error.}



\begin{table}
\caption{\NEW{RMSE comparison between Li et al.~\shortcite{Li17} and our method. Due to the use of different parametrizations, we cannot compute RMSE on specular terms for Li et al.~\shortcite{Li17}. As their output albedo maps can have a different scaling than the ground truth with respect to lighting, we evaluate the re-rendering and diffuse albedo RMSE with multiple scaling factors on the albedo, and keep the best one (0.27).}}
\vspace{-2mm}
\begin{tabular}{|c|c|c|}
  \hline 
  Method & Li et al. & Ours \\ 
  \hline 
  Re-Rendering error & 0.169 & 0.083 \\ 
  \hline 
  Normal error & 0.046 & 0.035
   \\ 
  \hline 
  Diffuse albedo error & 0.090 & 0.019 \\ 
  \hline 
  Specular albedo error & NA & 0.050 \\ 
  \hline 
  Specular roughness error & NA & 0.129 \\ 
  \hline 
\end{tabular}   
\label{table:quantitativeComparison}
\vspace{-1mm}
\end{table}

Overall, our method reproduces the specularity of the ground truth more accurately, as evidenced by the sharpness of reflections and highlights in the re-renderings. We believe this is due to our use of near-field flash illumination, as the apparent size and intensity of the highlight caused by the flash is strongly indicative of the overall glossiness and albedo levels. The method of Li et al.~\shortcite{Li17} must rely on more indirect and ambiguous cues to make these inferences. While such cues are available in the input images -- for example, the reflections of the illumination environment are blurred to different degrees -- their method has not reached an equally accurate estimate of the specular roughness.

Similarly, flash illumination highlights the surface normal variations by introducing spatially varying directional shading effects into the image. Such variations do also have a characteristic appearance in environment-lit images, but interpreting these cues may be more difficult due to ambiguities and uncertainties related to the unknown lighting environment. Consequently, the normal maps recovered by Li et al.~\shortcite{Li17} are also less accurate than ours.

\begin{figure*}[t]
 \begin{tabular}{ccccccc}

\hspace{-9mm} \begin{sideways} \hspace{-8mm}  \small{\cite{Li17}} \end{sideways} &
\hspace{-6mm} \includegraphics[align=c, width=0.14\linewidth]{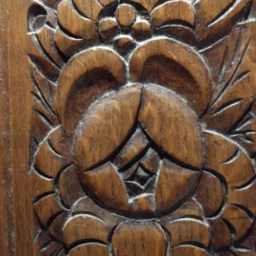} & 
 \includegraphics[align=c, width=0.14\linewidth]{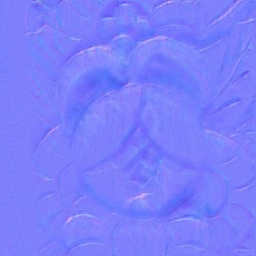} & 
\hspace{-3mm} \includegraphics[align=c, width=0.14\linewidth]{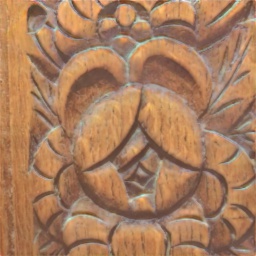} & 
\hspace{-3mm} \includegraphics[align=c, width=0.14\linewidth]{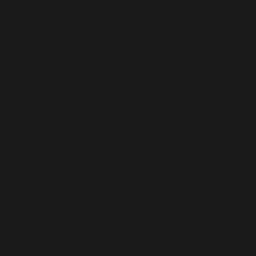} & 
\hspace{-3mm} \includegraphics[align=c, width=0.14\linewidth]{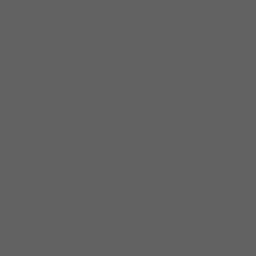} & 
 \includegraphics[align=c, width=0.14\linewidth]{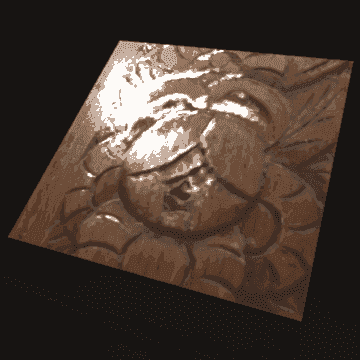} \\
\vspace{-3mm} \\ 

\hspace{-9mm} \begin{sideways} \hspace{-8mm}  \small{Our approach} \end{sideways} &
\hspace{-6mm} \includegraphics[align=c, width=0.14\linewidth]{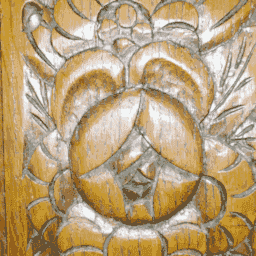} & 
 \includegraphics[align=c, width=0.14\linewidth]{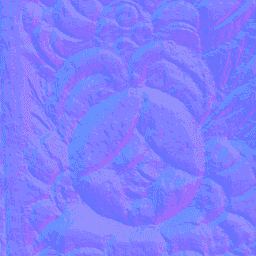} & 
\hspace{-3mm} \includegraphics[align=c, width=0.14\linewidth]{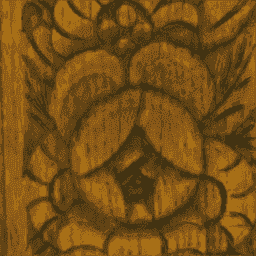} & 
\hspace{-3mm} \includegraphics[align=c, width=0.14\linewidth]{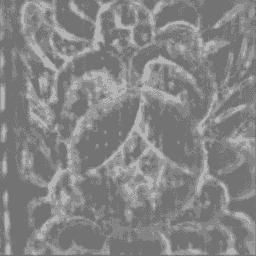} & 
\hspace{-3mm} \includegraphics[align=c, width=0.14\linewidth]{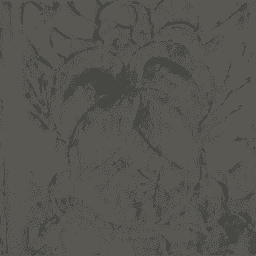} & 
 \includegraphics[align=c, width=0.14\linewidth]{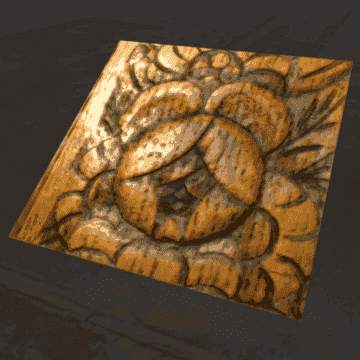} \\
\\

\hspace{-9mm} \begin{sideways} \hspace{-8mm}  \small{\cite{Li17}} \end{sideways} &
\hspace{-6mm} \includegraphics[align=c, width=0.14\linewidth]{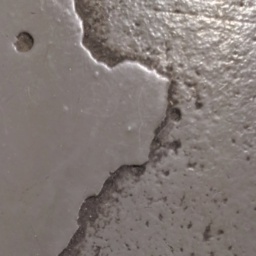} & 
 \includegraphics[align=c, width=0.14\linewidth]{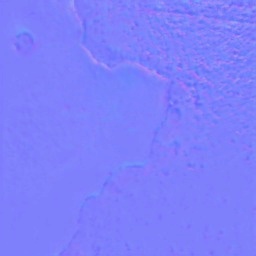} & 
\hspace{-3mm} \includegraphics[align=c, width=0.14\linewidth]{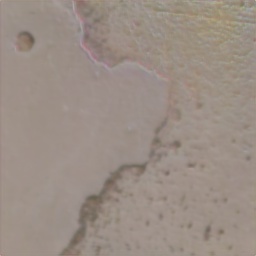} & 
\hspace{-3mm} \includegraphics[align=c, width=0.14\linewidth]{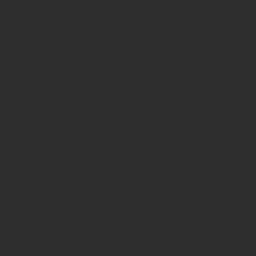} & 
\hspace{-3mm} \includegraphics[align=c, width=0.14\linewidth]{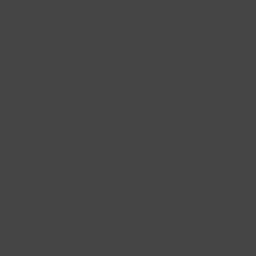} & 
 \includegraphics[align=c, width=0.14\linewidth]{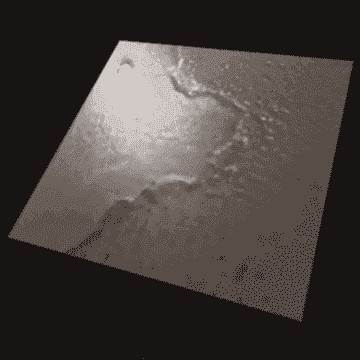} \\

\vspace{-3mm} \\ 

\hspace{-9mm} \begin{sideways} \hspace{-8mm}  \small{Our approach} \end{sideways} &
\hspace{-6mm} \includegraphics[align=c, width=0.14\linewidth]{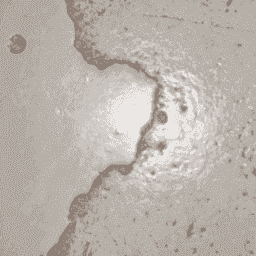} & 
 \includegraphics[align=c, width=0.14\linewidth]{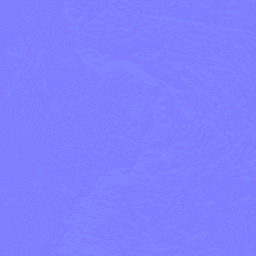} & 
\hspace{-3mm} \includegraphics[align=c, width=0.14\linewidth]{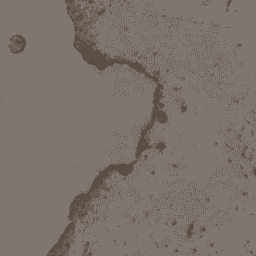} & 
\hspace{-3mm} \includegraphics[align=c, width=0.14\linewidth]{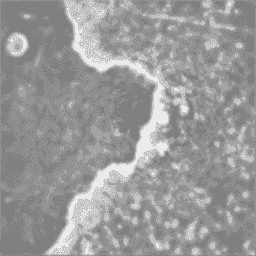} & 
\hspace{-3mm} \includegraphics[align=c, width=0.14\linewidth]{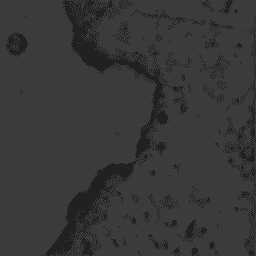} & 
 \includegraphics[align=c, width=0.14\linewidth]{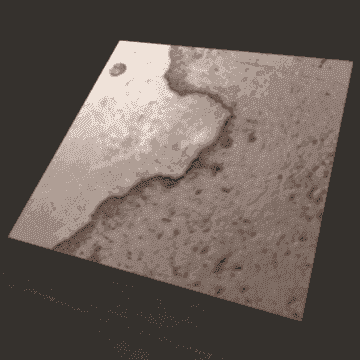} \\

&\hspace{-6mm}\small{Input}&\small{Normal}&\hspace{-3mm}\small{Diffuse albedo}&\hspace{-3mm}\small{Roughness}&\hspace{-3mm}\small{Specular albedo}&\small{Re-rendering}

\end{tabular}

\vspace{-2mm}
\caption{Comparison with Li et al. \protect \shortcite{Li17} on real-world data. We captured the input photographs under flash lighting for our method and under environment lighting for the method by Li et al., in agreement with the type of input assumed by each method. Please refer to Figure~\ref{fig:SyntheticComparisonMicrosoft} for notes on interpreting the results.}

\label{fig:ComparisonMicrosoftRealData}
\end{figure*}

We then compare the two methods on real pictures, captured with a flash for our approach and without for the approach by Li et al. (Figure~\ref{fig:ComparisonMicrosoftRealData}). Overall, the relative performance of the methods appears similar to the synthetic case.

\begin{figure*}

\begin{tabular}{cccccc}

 \includegraphics[align=c, width=0.15\textwidth]{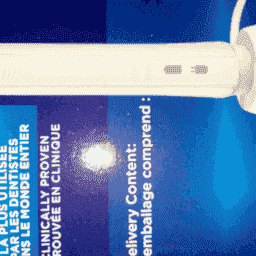} &
 \includegraphics[align=c, width=0.15\textwidth]{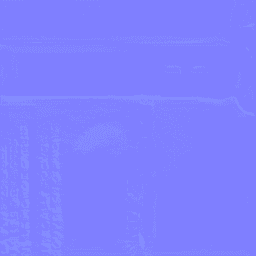} &
 \hspace{-3mm} \includegraphics[align=c, width=0.15\textwidth]{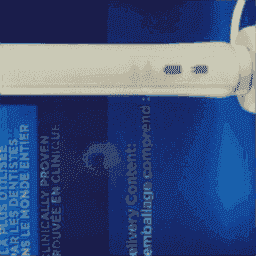} &
 \hspace{-3mm} \includegraphics[align=c, width=0.15\textwidth]{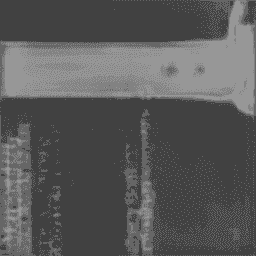} &
 \hspace{-3mm} \includegraphics[align=c, width=0.15\textwidth]{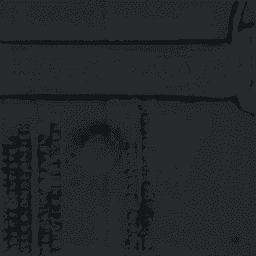} &
 \includegraphics[align=c, width=0.15\textwidth]{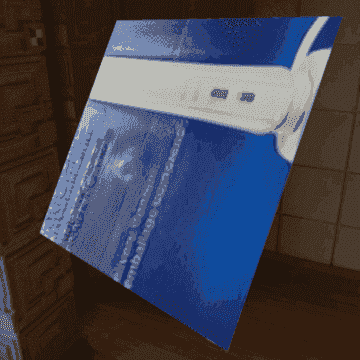} \\\\

 \includegraphics[align=c, width=0.15\textwidth]{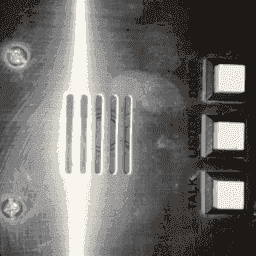} &
 \includegraphics[align=c, width=0.15\textwidth]{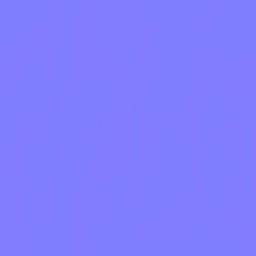} &
 \hspace{-3mm} \includegraphics[align=c, width=0.15\textwidth]{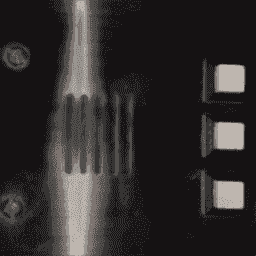} &
 \hspace{-3mm} \includegraphics[align=c, width=0.15\textwidth]{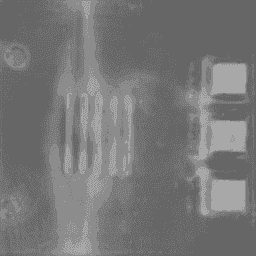} &
 \hspace{-3mm} \includegraphics[align=c, width=0.15\textwidth]{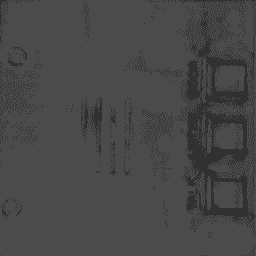} &
 \includegraphics[align=c, width=0.15\textwidth]{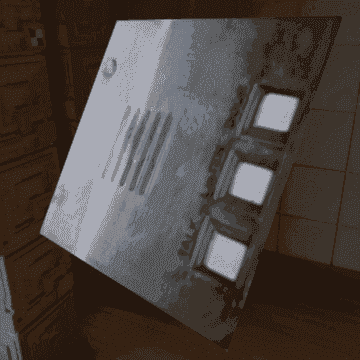} \\\\
 
 \includegraphics[align=c, width=0.15\textwidth]{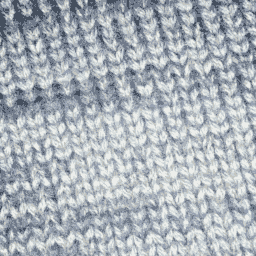} &
 \includegraphics[align=c, width=0.15\textwidth]{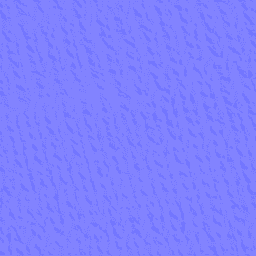} &
 \hspace{-3mm} \includegraphics[align=c, width=0.15\textwidth]{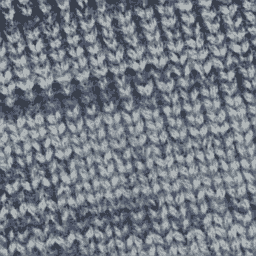} &
 \hspace{-3mm} \includegraphics[align=c, width=0.15\textwidth]{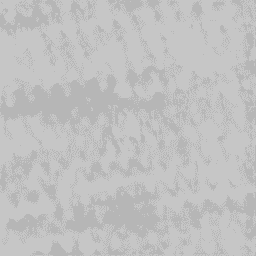} &
 \hspace{-3mm} \includegraphics[align=c, width=0.15\textwidth]{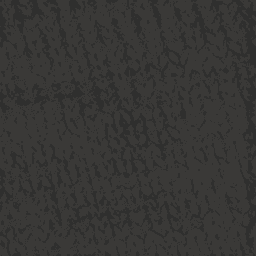} &
 \includegraphics[align=c, width=0.15\textwidth]{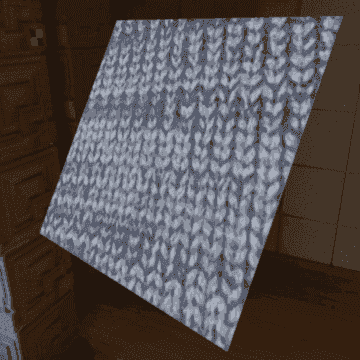} \\
 \small{Input} & \small{Normal} & \hspace{-3mm}\small{Diffuse} & \hspace{-3mm}\small{Roughness} & \hspace{-3mm}\small{Specular} & \small{Re-rendering}
 \end{tabular}
 
 \vspace{-2mm}
	\caption{Failure cases and performance on materials violating \REM{the}\NEW{our} assumptions. Our method generally struggles with otherwise uniform surfaces exhibiting structured albedo detail, such as the text and the photograph on the product packaging (top). Highly concentrated specular highlights are sometimes missed and result in overestimated roughness and occasional highlight removal artifacts (top). Materials outside the scope of the training data (e.g. anisotropic brushed metal, center) cannot be reproduced properly, and result in an undefined assignment of the apparent shading effects (the streak of the specular highlight) into the various maps of the SVBRDF. Nevertheless, the method \NEW{can} produce reasonable approximations for materials violating the assumptions, with varying degrees of success, as seen on the fuzzy wool (bottom).}

\label{fig:FailureCases}

\end{figure*}

\subsection{Limitations}
Despite the diversity of results shown, the architecture of our deep network imposes some limitations on the type of images and materials we can handle.
 
In terms of input, our network processes images of $256\times256$ pixels, which prevents it from recovering very fine details. While increasing the resolution of the input is an option, it would increase the memory consumption of the network and may hinder its convergence.
Recent work on iterative, coarse-to-fine neural image synthesis represents a promising direction to scale our approach to high-resolution inputs \cite{ChenKoltun2017,Karras17}.
Our network is also limited by the low dynamic range of input images. In particular, sharp, saturated highlights sometimes produce residual artifacts in the predicted maps as the network struggles to inpaint them with plausible patterns (Figure~\ref{fig:FailureCases}). \NEW{We also noticed that our network tends to produce correlated structures in the different maps. As a result, it fails on materials like the one in Figure~\ref{fig:FailureCases} (top row), where the packaging has a clear coat on top of a textured diffuse material. This behavior may be due to the fact that most of the artist-designed materials we used for training exhibit correlated maps.} 
\NEW{Finally, while our diverse results show that our network is capable of exploiting subtle shading cues to infer SVBRDFs, we observed that it resorts to naive heuristics in the absence of such cues. For example, the normal map for the wool knitting in Figure~\ref{fig:FailureCases} suggests a simple ``dark is deep'' prior.}


In terms of output, our network parameterizes an SVBRDF with four maps. Additional maps should be added to handle a wider range of effects, such as anisotropic specular reflections. 
The Cook-Torrance BRDF model we use is also not suitable for materials like thick fabric or skin, which are dominated by multiple scattering. Extending our approach to such materials would require a parametric model of their spatially-varying appearance, as well as a fast renderer to compute the loss.
\NEW{Finally, since our method only takes a fronto-parallel picture as input, it never observes the material sample at grazing angle, and as such cannot recover accurate Fresnel effects.}

\vspace{-0.2cm}
\section{Conclusion}
The casual capture of realistic material appearance is a critical challenge of 3D authoring. We have shown that a neural network can reconstruct complex spatially varying BRDFs given a single input photograph, and based on training from synthetic data alone. In addition to the quantity and realism of our training data, the quality of our results stems from an approach that is both aware of how SVBRDF maps interact together -- thanks to our rendering loss -- and capable of fusing distant information in the image -- thanks to its global feature track. Our method generalizes well to real input photographs and we show that a single network can be trained to handle a large variety of materials.

\section*{Acknowledgments}
\NEW{We thank the reviewers for numerous suggestions on how to improve the exposition and evaluation of this work. 
We also thank the Optis team, V. Hourdin, A. Jouanin, M. Civita, D. Mettetal and N. Dalmasso for regular feedback and suggestions, 
S. Rodriguez for insightful discussions, 
Li et al.~\shortcite{Li17} and Weinmann et al.~\shortcite{weinmann2014} for making their code and data available, and J. Riviere for help with evaluation. This work was partly funded 
by an ANRT (http://www.anrt.asso.fr/en) CIFRE scholarship between Inria and Optis, by the Toyota Research Institute and EU H2020 project 727188 EMOTIVE, and by software and hardware donations from Adobe and Nvidia. Finally, we thank Allegorithmic and Optis for facilitating distribution of our training data and source code for non-commercial research purposes, and all the contributors of Allegorithmic Substance Share.}

\bibliographystyle{ACM-Reference-Format}
\bibliography{bibtex/bibliography} 
\end{document}